\documentclass[%
floatfix, reprint,
 amsmath,amssymb,
 aps,
prc,
]{revtex4-2}
\usepackage{amsmath}
\allowdisplaybreaks[4]
\usepackage{amssymb}
\usepackage{graphicx}% Include figure files
\usepackage{dcolumn}% Align table columns on decimal point
\usepackage{bm}
\usepackage{comment}
\usepackage{tikz}
\usepackage{booktabs}
\usepackage{soul, color, xcolor}
\usepackage{ulem}

\usepackage[colorlinks=true,linkcolor=blue,citecolor=blue,urlcolor=blue]{hyperref}
\soulregister\cite7
\soulregister\eqref7
\usetikzlibrary{decorations.markings}
\usetikzlibrary{decorations.pathmorphing, decorations.markings, arrows}
\tikzset{
  arrow at/.style={
    decoration={
      markings,
      mark=at position #1 with {\arrow{latex}}
    },
    postaction={decorate}
  }
}% bold math

\begin{document}

\preprint{PS/123-QAED}

\title{$0\nu\beta\beta$ decay nuclear matrix elements under Left-Right symmetric model from the spherical quasi-particle random phase approximation method with realistic force}% Force line breaks with \\

\author{Ri-Guang Huang$^{1,2}$}~\email{huangriguang@impcas.ac.cn}
%\affiliation{Institute of Modern Physics, Chinese Academy of sciences, Lanzhou, 730000, China}
\author{You-Cai Chen$^{3}$}~\email{chenyc20@mails.jlu.edu.cn}

\author{Dong-Liang Fang$^{1,2}$}~\email{dlfang@impcas.ac.cn}
\affiliation{Institute of Modern Physics, Chinese Academy of sciences, Lanzhou, 730000, China$^{1}$}
\affiliation{School of Nuclear Sciences and Technology, University of Chinese Academy of Sciences, Beijing 101408, China$^{2}$}
\affiliation{Jilin University, Changchun 130012, People's Republic of China$^{3}$}


\begin{abstract}
We perform the calculation of nuclear matrix elements for the neutrinoless double beta decays under a Left-Right symmetric model mediated by light neutrino, and we adopt the spherical quasi-particle random-phase approximation (QRPA) approach with realistic force. For eight nuclei: $^{76}$Ge, $^{82}$Se, $^{96}$Zr, $^{100}$Mo, $^{116}$Cd, $^{128}$Te, $^{130}$Te and $^{136}$Xe, related nuclear matrix elements are given.  We analyze each term and the details of contributions of different parts are also given.
For the $q$ term, we find that the weak-magnetism components of the nucleon current contribute equally as other components such as axial-vector. 
We also discuss the influence of short-range correlations on these NMEs. It is found that $R$ term are more sensitive to the short range correlation than other terms due to the large portion of the contribution from high exchange momenta.   
\end{abstract}

\maketitle

%\tableofcontents
\section{Introduction}
As a rare and unobserved process, the neutrinoless double beta ($0\nu\beta\beta$) decay attracts attentions from communities of nuclear and particle physics.
This second-order weak process, with an even-even nucleus decaying to its neighboring with two additional protons and with two electrons emitted, has no neutrinos in its products. This is different from the other mode of double beta decay, namely two neutrinos double beta decay which has been observed for years.
The absence of neutrinos in the decay product indicates one thing, that this process is a lepton-number-violating ($\Delta L=2$) process, and its discovery suggesting the new physics beyond the particle standard model. 
  
Neutrino is the light Fermions ever known in our universe. It is still surrounded by many unsolved puzzles, including the origin of its mass, its mass hierarchy, and whether it is Dirac or Majorana particle \cite{Schechter1982}.
The $0\nu\beta\beta$ decay is the key for our understanding of these problems.
Accurately calculated nuclear matrix elements(NMEs) are particularly important in the search for $0\nu\beta\beta$ decay, as the half-lives of this process are dependent on its NMEs, which describe the nuclear transition during this process. 
It is the key to extract the so-called effective neutrino mass $|m_{\beta\beta}|$ or other new physics parameters from measured half life of $0\nu\beta\beta$ depending on the under-lying mechanisms, once the process is observed. 
There are a large amount of nuclear calculations with different many-body approaches focusing on the determination of NMEs, for instance, the works from large scale shell model (LSSM) \cite{Menendez2018,Horoi2016,Senkov2013}, the quasi-particle
random-phase approximation (QRPA) \cite{Simkovic1999,Rodin2006,Kortelainen2007,Simkovic2013,Suhonen-2015,fang2018,Lv2023},
the interacting boson model (IBM) \cite{Barea2015,Deppisch2020} and energy-density functional (EDF) theory \cite{Rodriguez2010,Song2014,Song2017,Yao2015}.
However, a significant challenge remains: NMEs derived from different nuclear models can differ by as much as a factor of three \cite{Yao2022,Matteo2023}. 
%Many researchers are striving to enhance the precision of these calculations and reduce uncertainty.
A promising avenue of reducing the discrepancy in calculations of NMEs is to develop the \textit{ab initio} methods, which is still challenging task especially for heavier nuclei \cite{Yao2020}.

Most of the above calculations focus on the neutrino mass mechanism where the only new physics ingredient one introduce is the neutrino mass term. These mass terms are most probably come from the so-call neutrino see-saw mechanism, they are usually embedded in much more complicated new physics model. One of such model which could naturally incorporate such mechanism is an extension to the standard model, the Left-Right symmetric model (LRSM). Within the framework of LR symmetric model, besides the usual neutrino mass mechanism, one has more underlying mechanisms for $0\nu\beta\beta$-decay from the $q$ part of the neutrino propagator besides the mass part. The $q$ term of neutrino propagator generally induces the $\eta$ mechanism associated with the mixing of left- and right-handed gauge bosons and $\lambda$ mechanism related to the mass ratio of these two bosons.  

For these extra NMEs of LRSM, there have been less studies and inadequate nuclear many-body calculations compared to that of the mass mechanism. The existing limited calculations adopt different many-body approaches, including LSSM \cite{NSM-PRL}, QRPA \cite{Muto1989,Pantis1996,SUHONEN1998}, and projected Hatree-Fock Boglyubov(PHFB) \cite{Tomoda1991}. 
More recently, several new calculations of NMEs relevant to LRSM have been conducted.
These include evaluations of the NMEs associated with the mass and $\lambda$ mechanisms, using LSSM \cite{Sarkar2020,Iwata2021} and  QRPA \cite{Simkovic2017} approaches, which incorporate higher-order nucleon currents based on the improved formalism from Ref. \cite{Reexam-LR}. 
It was found that the including pseudoscalar contributions enhance $qGT$ matrix elements significantly in the Ref. \cite{Sarkar2020,Iwata2021}.
Additionally, the competition of contributions in the decay rate between the mass and $\lambda$ mechanisms was also discussed in the QRPA calculations \cite{Simkovic2017} based on relevant calculated NMEs.
%The weak-magnetism currents is treated as a NLO contribution \cite{Cirigliano2017}.
Then the MM components in the $q$ terms have often been considered suppressed and hence neglected in the previous calculations.
However, recent work using the LSSM approach \cite{fang2024} have shown that these MM components are essential for the calculations of relevant LRSM NMEs.  
Till now, such QRPA calculations that incorporate these MM components in LRSM NMEs are still absent. 

Generally speaking, we still lack thorough investigation over the LRSM NMEs from various calculations, and the calculations for these NMEs are far from abundant. Therefore, in this work, we calculate the NMEs for the LRSM using the pn-QRPA approach, employing a large model space and realistic NN interactions, for all the mechanisms mediated by light neutrinos, the mass, $\eta$ and $\lambda$ mechanisms.
%Our calculations also incorporate several components in relevant NMEs to figure out whether their contributions are significant.
%In addition, we also evaluate the $R$ and $P$ terms associated with $\eta$ mechanism, which will be beneficial for future discussions on the contributions of this mechanism and the constrains of corresponding lepton number violating parameters.

This article is structured as follows: we begin with a brief introduction to the underlying formalism of the LRSM and the formalism of the QRPA method.
Next, we present the results of the NMEs and discuss the various contributions of these terms. Furthermore, we will also examine the influence of different SRC parametrizations. Finally, we draw our conclusions.

\section{formalism}
Ignoring the contribution from heavy neutrino, the 0$\nu\beta\beta$ half-life for the LR symmetric model with $\lambda$ and $\eta$ mechanisms can be written as \cite{Doi:1985dx,Reexam-LR, fang2024}:
%\begin{equation}
\begin{align}
	[T_{1/2}^{0\nu}]^{-1}=& \bigg\{
		 C_{mm}\left(\frac{|m_{\beta\beta}|}{m_e}\right)^2 +C_{\lambda\lambda}{\left<\lambda\right>}^2
	+C_{\eta\eta}\left<\eta\right>^2  \nonumber  \\
	&+C_{m\lambda}\frac{|m_{\beta\beta}|}{m_e}\left<\lambda\right>cos\psi_1 
	+C_{m\eta}\frac{|m_{\beta\beta}|}{m_e}\left<\eta\right>cos\psi_2  \nonumber \\
	&
	+C_{\lambda\eta}\left<\lambda\right>\left<\eta\right>cos(\psi_1 -\psi_2)   \bigg\}.
\end{align}
%\end{equation}
%\begin{equation}
%\begin{split}
%	[T_{1/2}^{0\nu}]^{-1}&=g_A ^4 |M_{GT}|^2 \\
%	&\times\bigg\{ 
%	{C}_{mm}\left(\frac{|m_{\beta\beta}|}{m_e}\right)^2  
%	+{C}_{m\lambda}\frac{|m_{\beta\beta}|}{m_e}\left<\lambda\right>cos\psi_1 \\
%	&\quad+{C}_{m\eta}\frac{|m_{\beta\beta}|}{m_e}\left<\eta\right>cos\psi_2   
%	+{C}_{\lambda\lambda}{\left<\lambda\right>}^2  \\
%	&\quad+{C}_{\eta\eta}\left<\eta\right>^2 
%	+{C}_{\lambda\eta}\left<\lambda\right>\left<\eta\right>cos(\psi_1 -\psi_2)   \bigg\}. 
%\end{split}
%\end{equation}
Here, the $m_{\beta\beta}=U_{ej}^2 m_j$ is the effective neutrino mass. And other parameters such as $\left< \lambda \right>$ or $\left< \eta \right>$ are given in \cite{Reexam-LR}: $\left< \lambda \right>=|\lambda\sum_{j}U_{ej}T_{ej}^* (g_V ^\prime /g_V)|$ and $\left< \eta \right>=|\eta\sum_{j}U_{ej}T_{ej}^*|$. The two phase angles are $\psi_1 =arg\left[\left(\sum_j m_j U^2 _{ej}\right) \left(\sum_j U_{ej}T_{ej}^* (g_V ^\prime /g_V) \right)^* \right]$ and $\psi_2 =arg\left[\left(\sum_j m_j U^2 _{ej}\right) \left(\sum_j U_{ej}T_{ej}^*\right)^* \right]$. The $U$ and $T$ are the constituents of a generalized 6$\times$6 Pontecorvo-Maki-Nakagawa-Sakata (PMNS) matrix \cite{Xing:2011ur}. 
It is typically assumed that $g_V \approx g_V ^\prime$, leading to $\psi_1 \approx \psi_2$.

The explicit forms of coefficients $C^\prime$s, which are the combinations of nuclear matrix elements (NMEs) and phase-space factors (PSFs), can also be found in Ref. \cite{fang2024}, with $G_{01,02,03}=\mathcal{G}_{01,02,03}$, $G_{04,05,06}=\mathcal{G}_{04,05,06}/(m_e R)$ and $G_{07,08,09,10,11}=\mathcal{G}_{07,08,09,010,011}/(m_e R)^2$ : 
% \begin{align}
%   \left<\lambda\right>&=\left|\frac{C^{(6)}_{VR}}{2V_{ud}}\right|,\ \ \left<\eta\right>=-\left|\frac{C^{(6)}_{VL}}{2V_{ud}}\right|\nonumber\\
%   \varphi_1&=\arg\left[-m_{\beta\beta}\left(C^{(6)}_{VR}\right)^*\right]\nonumber\\
%   \varphi_2&=\arg\left[m_{\beta\beta}\left(C^{(6)}_{VL}\right)^*\right]
% \end{align}
%Where the coefficients $C$'s as functions of NMEs and phase space factors are defined as: 
%Here the coefficients in the Equation (\ref{equaiton:half life}) can be classified according to the terms $\omega$, $\bm{q}$, $P$ and $R$ as  ($C_I$, $I\in (mm,\ m\eta,\ m\lambda,\ \lambda\eta,\ \lambda\lambda,\ \eta\eta)$)
\begin{align}
C_{mm}&=G_{01}M^2_{{m}} \nonumber\\
C_{m\lambda}&=-G_{03}M_{{m}} M_{\omega -}+G_{04}M_{{m}} M_{q+}\nonumber\\
C_{m\eta}&=G_{03}M_{{m}} M_{\omega +}-G_{04} M_{{m}}  M_{q-} -G_{05}M_{{m}} M_P\nonumber \\
        &\ \ \ + G_{06}M_{{m}} M_R \nonumber\\
C_{\lambda\lambda}&=G_{02}M^2_{\omega -}+G_{011}M^2_{q+} -2 G_{010} M_{\omega -} M_{q+} \nonumber\\
C_{\eta\eta}& =G_{02} M^2_{\omega +}+G_{011} M^2_{q-} -2G_{010} M_{\omega +} M_{q-} \nonumber\\
            &\ \ \  + G_{08} M^2_P  + G_{09} M^2_R - G_{07} M_P M_R \nonumber\\
C_{\lambda\eta}
              & =-2G_{02}M_{\omega -}M_{\omega +} -2 G_{011} M_{q+} M_{q-} \nonumber\\
            & \ \ \ +2G_{010} \left(M_{\omega -}M_{q-}+M_{\omega +}M_{q+}\right),
\label{equation:coefficients}
\end{align}
with $M_{m}$, $M_{\omega}$, $M_{q}$, $M_{R}$ and $M_{P}$ explicitly included. These NMEs may consist of different parts:
\begin{eqnarray}\label{nmeparts}
M_{m}&=&-M_F+M_{GT}+M_T \nonumber \\
M_{\omega \pm}&=&\pm M_{\omega F}+M_{\omega GT\pm}+M_{\omega T {\pm}} \nonumber \\
M_{q \pm}&=&\pm M_{q F}+M_{q GT\pm}+M_{q T {\pm}} \nonumber \\
M_{R}&=&M_{RGT}+M_{RT}.
\end{eqnarray}

These different parts of NMEs $M_I$ can be written in a general form:
\begin{align}
	\label{NEM_eq}
	M_I = \langle 0_f ^+||h_I (r ,r_+)\mathcal{O}_I ||0_i ^+\rangle,
\end{align}
where $I$'s are different parts in eq.\eqref{nmeparts}.
$|0_i ^+\rangle$ and $|0_f ^+\rangle$ represent the ground states of initial and final nuclei.
The distance $r$ is defined as $|\boldsymbol{r}_m -\boldsymbol{r}_n|$, and $r_+$ is given by $|(\boldsymbol{r}_m +\boldsymbol{r}_n)/2|$, where $\boldsymbol{r}_{m(n)}$ are the coordinates of the decaying nucleons, labeled as $m$ or $n$, respectively.

For the mass, $q$, $R$ and $P$ terms, the neutrino potentials $h_I (r ,r_+)$ in (\ref{NEM_eq}) are expressed in the form:
\begin{align}
	\label{nu-potent}
	h_I (r,r_+)=f_{src}^2(r) \frac{2R}{\pi}\int f_I (q,r,r_+)\frac{qdq}{q+E_m -(E_i +E_f)/2},
\end{align}
where $E_i$ ($E_f$) is the ground state energy of the initial (final) nucleus, and $E_m$ is the energy of the intermediate nucleus.
In order to account for the characteristics of nuclear force at short distance, which is missing in the nuclear wave functions from many-body calculations, a short-range correlation (SRC) function $f_{src}(r)=1-ce^{-ar^2}(1-br^2)$ \cite{Kortelainen2007-2,Gerald1976} is introduced in the calculations of neutrino potentials (\ref{nu-potent}).
In our work, we apply the Argonne V18 and CD-Bonn parametrizations of the SRC, where the parameters a, b, and c in the $f_{src}$ are taken from Ref. \cite{simkovic-SRC}. The effects of SRC on the NMEs will be discussed later. 
%A scheme dependent function $f_{src}(r)=1-ce^{-ar^2}(1-br^2)$ is introduced to account for the nucleon-nucleon short-range correlations(SRC)\cite{SRC_Suho,SRC_2}. In this calculation, two interactions Argonne V18 and CD-Bonn \cite{SRC-CDBonn} are applied to the SRC. 
With the inclusion of pseudo-scalar and weak-magnetism terms, as an usual convention, the $f^, s$ functions in the neutrino potentials are given by the following form\cite{fang2024}.  For the mass term:
%\begin{equation}
\begin{align}
\label{mass_terms}
	f_F=&j_0(qr) g_V ^2(q^2) \nonumber\\
	f_{GT}=&f_{GT}^{AA}+f_{GT}^{AP}+f_{GT}^{PP}+f_{GT}^{MM}\nonumber\\
	=&{j_0(qr)}\left(g_A ^2 (q^2)-\frac{g_A(q^2)g_P (q^2)}{m_N}\frac{q^2}{3} \right. \nonumber
	 \\
	&+ \left.\frac{g_P^2 (q^2)}{4m^2 _N}\frac{q^4}{3}+\frac{g_M ^2(q^2)}{4m_N ^2}\frac{2q^2}{3} \right)\\
	f_{T}=&f_{T}^{AP}+f_{T}^{PP}+f_{T}^{MM}\nonumber\\
	=&{j_2(qr)}\left(\frac{g_A(q^2)g_P(q^2)}{m_N}\frac{q^2}{3}-\frac{g_P^2(q^2)}{4m_N^2}\frac{q^4}{3} \right. \nonumber\\
	+& \left.\frac{g^2_M(q^2)}{4m_N^2}\frac{q^2}{3}\right) \nonumber,
\end{align}
%\end{equation}
for the $q$ term:
%\begin{equation}
\label{q_terms}
\begin{align}
	f_{qF}=&j_1(qr) qrg^2 _V(q^2)  \nonumber\\
	f_{qGT\pm}=&f_{qGT}^{AA}+f_{qGT}^{AP}+f_{qGT}^{PP}\mp f_{qGT}^{MM}  \nonumber\\
	=&\frac{1}{3}{j_1(qr)qr} \left(g_A^2(q^2)+\frac{g_A(q^2)g_P(q^2)q^2}{m_N} \right. \nonumber\\
	 &-\left.\frac{g_P^2(q^2)q^4}{4m_N^2}\mp \frac{g_M^2(q^2)q^2}{2m_N^2}\right) \nonumber\\   
	 f_{qT\pm}=&f_{qT}^{AA}+f_{qT}^{AP}+f_{qT}^{PP}\mp f_{qT}^{MM} \\
    =&\frac{2}{3}{j_1(qr)qr}\left(-g_A ^2(q^2) + g_A(q^2)g_P(q^2)\frac{q^2}{2m_N}\right) \nonumber\\
    &-\frac{q^3 r}{20m_N^2}g^2_P(q^2) q^2\left(2j_1(qr)/3+ j_3(qr)\right) \nonumber\\
    &\mp \frac{q^3 r}{30m_N^2}g^2_M(q^2) \left(j_1(qr)+3j_3(qr)\right), \nonumber
\end{align}	
%\end{equation}
and for the R and P terms:
\begin{equation}
\label{RG_equation}	
\begin{aligned}
	f_{RGT}& =\frac{-R}{3m_{N}}{g_{A}(q^{2})g_{M}(q^{2})}j_{0}(qr)q^{2}  \\
	f_{RT}& =\frac{-R}{6m_{N}}{g_{A}(q^{2})g_{M}(q^{2})}j_{2}(qr)q^{2}  \\
	f_{P}& ={g_{V}(q^{2})g_{A}(q^{2})}j_{1}(qr)q r_+, 
\end{aligned}
\end{equation}
where $R=1.2A^{1/3}$ is the nuclear radius, and $j_{\lambda}$ ($\lambda=0,1,2,3$) is the spherical Bessel function with rank $\lambda$.
Considering the effects of finite nucleon size, 
the vector, axial-vector, weak-magnetism and induced pseudo-scalar momentum dependent form factors in the $f_I$ function are considered: 
\begin{equation}
\begin{aligned}
		g_V(q^2)&=g_V/[1+q^2/(\Lambda_V)^2]^{2}\\
		g_A(q^2)&=g_A/[1+q^2/(\Lambda_A)^2]^{2}\\
		g_M(q^2)&=[(\mu_p-\mu_n)+1]g_V(q^2)\\
		g_P(q^2)&=\frac{2m_N g_A(q^2)}{q^2+m_\pi ^2}.\\
\end{aligned}
\end{equation}
We take $g_V=1.0$ and $g_A=1.27$ in this work. 
$m_\pi$ and $m_N$ are the pion and nucleon mass, respectively.  The anomalous nucleon magnetic moment is $\mu_p -\mu_n \approx 3.71$. The cutoffs are $\Lambda_V=0.85$GeV, $\Lambda_A=1.086$GeV, respectively. 

Note that we change the definition of $M^{0\nu}_q$, this is due to the fact that earlier calculation has only the axial-vector included, induced current such as Pseudo-Scalar and weak magnetism terms are neglected. If these terms are included, we find that a factor of $1/3$ should be absorbed in the definition of $M_{qGT}^{AA}$ in the sense that we treat $M_{q}$ on equal footing as $M_{m}$ or $M_{\omega}$. In general, $M_{qGT}=\frac{1}{3}M_{qGT}^{o}$, $M_{qF}=M_{qF}^{o}$ and $M_{qT}=-2M_{qT}^{o}$, where the NME parts with superscript $o$ refers to those defined in \cite{Doi:1985dx,Simkovic2017,fang2024}. In the meantime, the corresponding phase space factors are also changed by dividing a factor $1/3$.

For the neutrino potential $h_I (r)$ with $I=\omega F$, $\omega GT$ and $\omega T$ in $\omega$ terms,  the expressions have a little difference from those in (\ref{nu-potent}), 
\begin{align}
\label{omega}
	h_I (r)=f_{src}^2(r) \frac{2R}{\pi}\int f_I (qr)\frac{q^2 dq}{\left[q+E_n -(E_i +E_f)/2\right]^2}.
\end{align}
Here, the $f^{\prime} s$ functions of the $\omega$ term are:

\begin{align}
	\label{omega_terms}
	f_{\omega F}=&j_0(qr) g_V ^2(q^2) \nonumber\\
	f_{\omega GT\pm}=&f_{GT}^{AA}+f_{GT}^{AP}+f_{GT}^{PP}\pm f_{GT}^{MM}\nonumber\\
	=&j_0(qr)\left(g_A ^2 (q^2)-\frac{g_A(q^2)g_P (q^2)}{m_N}\frac{q^2}{3} \right. \nonumber
	\\
	&+ \left.\frac{g_P (q^2)}{4m^2 _N}\frac{q^4}{3}\pm\frac{g_M ^2(q^2)}{4m_N ^2}\frac{2q^2}{3} \right)\\
	f_{\omega T\pm}=&f_{T}^{AP}+f_{T}^{PP}\pm f_{T}^{MM}\nonumber\\
	=&j_2(qr)\left(\frac{g_A(q^2)g_P(q^2)}{m_N}\frac{q^2}{3}-\frac{g_P^2(q^2)}{4m_N^2}\frac{q^4}{3} \right. \nonumber\\
	\pm& \left.\frac{g^2_M(q^2)}{4m_N^2}\frac{q^2}{3}\right) \nonumber.
\end{align}

Meanwhile the operator $\mathcal{O}_I$ in the NMEs can be expressed as:
\begin{equation}
\begin{aligned}
	\mathcal{O}_{F,qF,\omega F}&=1\\
	\mathcal{O}_{GT,qGT,\omega GT,RG}&=\boldsymbol{\sigma}_m \cdot \boldsymbol{\sigma}_n \\
	\mathcal{O}_{T,qT,\omega T,RT}&=3(\boldsymbol{\sigma}_m \cdot \hat{\boldsymbol{r}}) (\boldsymbol{\sigma}_n \cdot \hat{\boldsymbol{r}}) -\boldsymbol{\sigma}_m \cdot \boldsymbol{\sigma}_n \\
	\mathcal{O}_{P}&=i(\boldsymbol{\sigma}_m - \boldsymbol{\sigma}_n) \cdot (\hat{\boldsymbol{r}} \times \hat{\boldsymbol{r}}_+),\\
\end{aligned}
\end{equation}
where $\hat{\boldsymbol{r}}=\boldsymbol{r}/r$ and $\hat{\boldsymbol{r}}_+=\boldsymbol{r}_+/r_+$. 

To calculate these NMEs, certain many-body approaches are needed. In this work, the spherical proton-neutron quasi-particle random-phase approximation(pn-QRPA) with realistic force is employed. Although the QRPA approach lacks higher-order many-body correlations, it can incorporate a larger model space in the calculations compared to the nuclear shell model.
Within the QRPA framework, we choose the BCS vacua as the initial and final ground states of the parent and daughter nucleus, while the virtual states of  the intermediate odd-odd nucleus can be expressed as one phonon excitations:
\begin{equation}
|J^\pi m\rangle = Q^{J^\pi\dagger}_m |0\rangle={\sum_{pn}}(X^{J^\pi}_{m,pn} A^\dagger_{pn}+Y^{J^\pi}_{m,pn}\tilde{A}_{pn})|0\rangle .
\end{equation}
Here $m$ refers to the index of the intermediate states and $A_{pn}^\dagger=\alpha_p^\dagger \alpha_n^\dagger$ is the two quasi-particle creation operator, with $\alpha_\tau$ the quasi particle and $|0\rangle$ the vacuum state approximated as BCS vacuum \cite{Simkovic1999}.

The forward $X$ and backward $Y$ amplitudes are obtained by solving the following equations:  
\begin{align}
\begin{pmatrix}
	A&B\\
	-B&-A
\end{pmatrix}
\begin{pmatrix}
	X\\
	Y
\end{pmatrix}=\omega
\begin{pmatrix}X\\Y\end{pmatrix}.
\end{align}
These amplitudes are then the key inputs for the calculation of NMEs. The $\omega$ represents corresponding energy eigenvalues. The matrices $A$ and $B$ are defined as follows:
%\begin{equation}
\begin{align}
	&A_{pn,p^{\prime}n^{\prime}}^{J^{\pi}} =\delta_{pp^{\prime}}\delta_{nn^{\prime}}(E_p+E_n) \nonumber \\
	&-2g_{ph}(u_pv_nu_{p^{\prime}}v_{n^{\prime}}+v_pu_nv_{p^{\prime}}u_{n^{\prime}})F(pnp^{\prime}n^{\prime},J) \nonumber\\
	&-2(u_pu_nu_{p^\prime}u_{n^\prime}+v_pv_nv_{p^\prime}v_{n^\prime}) \nonumber\\
	&\times\left(g_{pp}^{T=1}G(pnp^{\prime}n^{\prime},JT=1)+g_{pp}^{T=0}G(pnp^{\prime}n^{\prime},J{T=0})\right),
\end{align}
%\end{equation}
%\begin{equation}
\begin{align}
	&B_{pn,p^{\prime}n^{\prime}}^{J^{\pi}} =-2g_{ph}(u_pv_nv_{p^{\prime}}u_{n^{\prime}}+v_pu_nu_{p^{\prime}}v_{n^{\prime}})F(pnp^{\prime}n^{\prime},J) \nonumber\\
	 &+2(u_pu_nv_{p^{\prime}}v_{n^{\prime}}+v_p v_n u_{p^{\prime}}u_{n^{\prime}}) \nonumber\\
	&\times\left(g_{pp}^{T=1}G(pnp^{\prime}n^{\prime},J{T=1})+g_{pp}^{T=0}G({pnp^{\prime}n^{\prime}},J{T=0})\right),
\end{align}
%\end{equation}
where $J^{\pi}$ denotes angular momentum and parity of virtual intermediate states.
The $E_p$ and $E_n$ are quasi-particle energy of proton and neutron respectively. 
They are obtained from sloving the BCS equations \cite{Simkovic1999}. $G$ and $F$ are G-matrix elements obtained from the Br\"uckner G-matrix equations. 

The parameters $g_{pp}$'s and $g_{ph}$ are the renormalized strengths for the two-body residual interaction of particle-particle and particle-hole channels, respectively. The $g_{ph}$ is traditionally fixed to reproduce the experimental position of the giant Gamow-Teller resonance (GTGR) strength, and $g_{ph}=1$ is used in current calculation. 
While for the particle-particle channel, there are two kinds of interaction matrix elements: isoscalar(T=0) and isovector(T=1). Following the procedure in the work \cite{Simkovic2013}, we employ  two parameters $g_{pp}^{T=0}$ and $g_{pp}^{T=1}$ to renormalize these two interaction strengths, which should be adjusted independently.
These two parameters are typically fixed in the $2\nu\beta\beta$ calculations.  
The isoscalar parameter $g_{pp}^{T=0}$ is determined by fitting the measured half-life of $2\nu\beta\beta$ decay. Meanwhile, the isovector parameter $g_{pp}^{T=1}$ is adjusted so that the Fermi NME $M^{2\nu}_F$ vanishes, thereby restoring the isospin symmetry in $2\nu\beta\beta$ \cite{Simkovic2013,Rodin2011}.

With the solutions of pn-QRPA equations and summing over all the intermediate states, the NMEs of $0\nu\beta\beta$ decay can be expressed as:

\begin{equation}
	\begin{aligned}
		M_{I}^{0\nu} =&\sum_{J^\pi,k_{i},k_{f},\mathcal{J}}\sum_{pp^{\prime}nn^{\prime}}(-1)^{j_{n}+j_{p^{\prime}}+J+\mathcal{J}}\sqrt{2\mathcal{J}+1} \nonumber \\
			&\times\begin{Bmatrix}j_p&j_n&J\\j_{n^{\prime}}&j_{p^{\prime}}&\mathcal{J}\end{Bmatrix}(pp^{\prime};\mathcal{J}|| h_I (r ,r_+) \mathcal{O}_I||nn^{\prime};\mathcal{J})\nonumber \\
			&\times\left(0_{f}^{+}||[c_{p^{\prime}}^{\dagger}\tilde{c}_{n^{\prime}}]_{J}||J_{k_{f}}^{\pi}\right)\langle J_{k_{f}}^{\pi}|J_{k_{i}}^{\pi}\rangle\nonumber \\
			& \times\left(J_{k_{i}}^{\pi}||[c_{p}^{\dagger}\tilde{c}_{n}]_{J}||0_{i}^{+}\right),
		\end{aligned} 
	\end{equation} 
where $k_i$ and $k_f$ label the different pnQRPA solutions corresponding to initial and final even-even nucleus.  
The one-body transition densities in the $M_I ^{0\nu} $ can be expressed as:
\begin{equation}
\begin{aligned}
\frac{(0_f^+||[c_{p^{\prime}}^\dagger\tilde{c}_{n^{\prime}}]_J||J_{k_f}^\pi)}{\sqrt{2J+1}}&=\big[v_{p^{\prime}}^{(f)}u_{n^{\prime}}^{(f)}X_{p^{\prime}n^{\prime}}^{J^\pi k_f}+u_{p^{\prime}}^{(f)}v_{n^{\prime}}^{(f)}Y_{p^{\prime}n^{\prime}}^{J^\pi k_f}\big]\\
\frac{(J_{k_{i}}^{\pi}||[c_{p}^{\dagger}\tilde{c}_{n}]_{J}||0_{i}^{+})}{\sqrt{2J+1}}&=\big[u_{p}^{(i)}v_{n}^{(i)}X_{pn}^{J^{\pi}k_{i}}+v_{p}^{(i)}u_{n}^{(i)}Y_{pn}^{J^{\pi}k_{i}}\big],
\end{aligned}
\end{equation} 
where the BCS occupation and vacancy amplitudes $v^{(i)}(v^{(f)})$, $u^{(i)}(u^{(f)})$ are derived from the solutions of BCS equations for the initial (final) even-even nucleus.  
These amplitudes contain important pairing information relevant to this nuclear process.
The overlap of the intermediate states between initial nuclei and final nuclei can be expressed as:
\begin{equation}
	\begin{aligned}\langle J^\pi _{k_f}|J^\pi _{k_i}\rangle=&\sum_{pn}\left(X_{pn}^{J^{\pi} k_f}X_{pn}^{J^{\pi} k_i}-Y_{pn}^{J^{\pi} k_f}Y_{pn}^{J^{\pi} k_i}\right)\\
    &\times\left(u_{p}^{(f)}u_{p}^{(i)}+v_{p}^{(f)}v_{p}^{(i)} \right)
    \left(u_{n}^{(f)}u_{n}^{(i)}+v_{n}^{(f)}v_{n}^{(i)}\right)\\
    &\times\langle BCS_f|BCS_i\rangle,
	\end{aligned} 
\end{equation}
where $\langle BCS_f|BCS_i\rangle$ is the overlap factor between initial and final BCS vacua, which is usually set to unity.

\section{RESULTS AND DISCUSSION}

\begin{table}[htb]
	\caption{\label{gpair} The proton and neutron paring strength $g_{pair}$ for the initial and final nuclei, as well as their average value $\left<g_{pair} \right>$. The isovector and isoscalar particle-particle parameters $g_{pp}^{T=1}$, $g_{pp}^{T=0}$ for the residual interaction. }
	\renewcommand{\arraystretch}{1.4}
%	\resizebox{1.0\linewidth}{!}{
\begin{tabular}{llllllll}
	\hline
	\hline
	 &$g_{pair,i}^{p}$ &$g_{pair,i}^{n}$ &$g_{pair,f}^{p}$ &$g_{pair,f}^{n}$ &$\left< g_{pair}\right>$
	 &$g_{pp}^{T=1}$   &$g_{pp}^{T=0}$ \\
	 \hline
$^{76}$Ge  & 0.84 & 0.94 & 0.87 & 0.97 & 0.91 & 0.92 & 0.59 \\
$^{82}$Se  & 0.78 & 0.94 & 0.83 & 0.99 & 0.89 & 0.91 & 0.60 \\
$^{96}$Zr  & 0.81 & 0.66 & 0.87 & 0.82 & 0.79 & 0.85 & 0.63 \\
$^{100}$Mo & 0.89 & 0.81 & 0.91 & 0.83 & 0.86 & 0.87 & 0.63 \\
$^{116}$Cd & 0.87 & 0.81 & 0.73 & 0.77 & 0.80 & 0.77 & 0.66 \\
$^{128}$Te & 0.77 & 0.87 & 0.81 & 0.87 & 0.83 & 0.86 & 0.57 \\
$^{130}$Te & 0.74 & 0.84 & 0.80 & 0.87 & 0.81 & 0.87 & 0.58 \\
$^{136}$Xe & 0.69 & 0.90 & 0.77 & 0.80 & 0.79 & 0.83 & 0.53 \\  
	\hline
	\hline
	\end{tabular}
            % }
\end{table}

	\begin{table*}[htb]

	\caption{\label{table-LR1} $0\nu\beta\beta$-decay NMEs for $^{76}$Ge, $^{82}$Se, $^{96}$Zr and $^{100}$Mo from pn-QRPA calculation. The results are obtained in the model spaces(N=0-7). For each nuclei, there are three kinds of NMEs presented in the table which are only different in the SRC, where AV18 (CD-Bonn) represents the results calculated with the Argonne (Charge-dependent-Bonn) type SRC parametrizations, w/o denotes the results without SRC.}
	\renewcommand{\arraystretch}{1.4}

			\begin{tabular}{c c c c c c c c c c c c c c}
				\hline    
				\hline  
			 	&&\multicolumn{3}{c}{$^{76}$Ge} &  \multicolumn{3}{c}{$^{82}$Se} &  \multicolumn{3}{c}{$^{96}$Zr} &  \multicolumn{3}{c}{$^{100}$Mo} \\
				\cmidrule(lr){3-5}\cmidrule(lr){6-8}\cmidrule(lr){9-11}\cmidrule(lr){12-14}
				%	\cmidrule(lr){13-14}\cmidrule(lr){15-16}\cmidrule(lr){17-18}				
				&&AV18 & cd-Bonn & w/o &AV18 & cd-Bonn & w/o &AV18 & cd-Bonn & w/o &AV18 & cd-Bonn & w/o  \\
				
				\hline
$M_{F}$         &                  & -1.482 & -1.600 & -1.522 & -1.360 & -1.463 & -1.390 & -1.033 & -1.108 & -1.057 & -1.952 & -2.090 & -1.994 \\
\cline{3-14}
$M_{GT}$        & AA               & 5.567  & 6.101  & 5.869  & 4.853  & 5.319  & 5.104  & 2.426  & 2.746  & 2.623  & 5.087  & 5.677  & 5.433  \\
& AP               & -2.126 & -2.362 & -2.342 & -1.890 & -2.096 & -2.075 & -1.165 & -1.306 & -1.299 & -2.226 & -2.487 & -2.468 \\
& PP               & 0.718  & 0.811  & 0.820  & 0.639  & 0.721  & 0.728  & 0.414  & 0.469  & 0.476  & 0.780  & 0.884  & 0.895  \\
& MM               & 0.819  & 1.000  & 1.092  & 0.725  & 0.882  & 0.963  & 0.479  & 0.586  & 0.642  & 0.897  & 1.096  & 1.200  \\
& total            & 4.667  & 5.169  & 5.024  & 4.051  & 4.491  & 4.353  & 1.971  & 2.272  & 2.198  & 4.197  & 4.753  & 4.604  \\
\cline{3-14}
$M_{T}$         & AP               & -0.989 & -0.986 & -0.960 & -0.931 & -0.928 & -0.905 & -0.743 & -0.742 & -0.723 & -1.296 & -1.293 & -1.261 \\
& PP               & 0.362  & 0.360  & 0.349  & 0.338  & 0.337  & 0.327  & 0.272  & 0.271  & 0.263  & 0.478  & 0.477  & 0.462  \\
& MM               & -0.239 & -0.239 & -0.228 & -0.221 & -0.221 & -0.212 & -0.178 & -0.178 & -0.169 & -0.311 & -0.311 & -0.296 \\
& total            & -0.775 & -0.774 & -0.752 & -0.730 & -0.728 & -0.709 & -0.582 & -0.581 & -0.564 & -1.011 & -1.009 & -0.982 \\
\hline
$M_{\omega F}$  &                  & -1.458 & -1.571 & -1.499 & -1.333 & -1.432 & -1.365 & -1.001 & -1.072 & -1.025 & -1.877 & -2.007 & -1.921 \\
\cline{3-14}
$M_{\omega GT}$ & AA               & 5.494  & 6.005  & 5.795  & 4.835  & 5.281  & 5.085  & 2.645  & 2.952  & 2.840  & 5.268  & 5.830  & 5.611  \\
& AP               & -2.095 & -2.327 & -2.309 & -1.867 & -2.070 & -2.050 & -1.167 & -1.305 & -1.299 & -2.207 & -2.462 & -2.446 \\
& PP               & 0.707  & 0.799  & 0.808  & 0.631  & 0.711  & 0.718  & 0.410  & 0.465  & 0.472  & 0.769  & 0.871  & 0.882  \\
& MM               & 0.805  & 0.983  & 1.076  & 0.714  & 0.870  & 0.950  & 0.473  & 0.580  & 0.636  & 0.883  & 1.080  & 1.183  \\
& $\omega_+$ total & 4.604  & 5.087  & 4.961  & 4.041  & 4.462  & 4.342  & 2.182  & 2.471  & 2.407  & 4.378  & 4.909  & 4.781  \\
& $\omega_-$ total & 3.607  & 3.868  & 3.627  & 3.156  & 3.383  & 3.164  & 1.595  & 1.752  & 1.619  & 3.283  & 3.570  & 3.315  \\
\cline{3-14}
$M_{\omega T}$  & AP               & -0.958 & -0.956 & -0.931 & -0.903 & -0.901 & -0.878 & -0.710 & -0.708 & -0.689 & -1.242 & -1.239 & -1.208 \\
& PP               & 0.351  & 0.350  & 0.339  & 0.329  & 0.328  & 0.318  & 0.260  & 0.260  & 0.252  & 0.459  & 0.458  & 0.444  \\
& MM               & -0.232 & -0.232 & -0.221 & -0.216 & -0.215 & -0.206 & -0.171 & -0.171 & -0.164 & -0.299 & -0.299 & -0.286 \\
& $\omega_+$ total & -0.752 & -0.750 & -0.729 & -0.708 & -0.706 & -0.688 & -0.555 & -0.554 & -0.538 & -0.969 & -0.967 & -0.941 \\
& $\omega_-$ total & -0.464 & -0.463 & -0.455 & -0.440 & -0.440 & -0.432 & -0.343 & -0.341 & -0.336 & -0.597 & -0.596 & -0.587 \\
\hline
$M_{qF}$        &                  & -0.944 & -0.971 & -0.857 & -0.886 & -0.910 & -0.806 & -0.704 & -0.722 & -0.647 & -1.355 & -1.387 & -1.247 \\
\cline{3-14}
$M_{qGT}$       & AA               & 1.419  & 1.484  & 1.325  & 1.217  & 1.273  & 1.130  & 0.465  & 0.504  & 0.415  & 1.129  & 1.201  & 1.030  \\
& AP               & 2.869  & 2.992  & 2.740  & 2.547  & 2.654  & 2.428  & 1.436  & 1.510  & 1.368  & 2.827  & 2.964  & 2.692  \\
& PP               & -1.271 & -1.329 & -1.226 & -1.136 & -1.186 & -1.094 & -0.693 & -0.728 & -0.670 & -1.330 & -1.395 & -1.284 \\
& MM               & -2.172 & -2.362 & -2.256 & -1.932 & -2.098 & -2.001 & -1.247 & -1.360 & -1.301 & -2.361 & -2.571 & -2.456 \\
& $q_+$ total      & 4.364  & 4.611  & 4.237  & 3.826  & 4.042  & 3.705  & 1.981  & 2.130  & 1.919  & 4.091  & 4.365  & 3.961  \\
& $q_-$ total      & 1.671  & 1.682  & 1.440  & 1.431  & 1.440  & 1.223  & 0.435  & 0.443  & 0.306  & 1.163  & 1.176  & 0.915  \\
\cline{3-14}
$M_{qT}$        & AA               & 3.510  & 3.504  & 3.414  & 3.306  & 3.300  & 3.219  & 2.590  & 2.585  & 2.520  & 4.393  & 4.385  & 4.273  \\
& AP               & -1.873 & -1.869 & -1.799 & -1.746 & -1.743 & -1.680 & -1.394 & -1.390 & -1.340 & -2.448 & -2.443 & -2.355 \\
& PP               & 0.561  & 0.560  & 0.532  & 0.517  & 0.515  & 0.490  & 0.414  & 0.413  & 0.393  & 0.726  & 0.724  & 0.689  \\
& MM               & 0.214  & 0.215  & 0.200  & 0.194  & 0.196  & 0.179  & 0.156  & 0.156  & 0.142  & 0.266  & 0.268  & 0.247  \\
& $q_+$ total      & 2.065  & 2.062  & 2.022  & 1.956  & 1.951  & 1.919  & 1.513  & 1.511  & 1.485  & 2.506  & 2.499  & 2.453  \\
& $q_-$ total      & 2.331  & 2.328  & 2.271  & 2.196  & 2.194  & 2.140  & 1.707  & 1.705  & 1.662  & 2.836  & 2.831  & 2.760  \\
\hline
& $RGT$             & 8.873  & 11.240 & 12.756 & 8.045  & 10.165 & 11.510 & 5.632  & 7.151  & 8.137  & 10.679 & 13.536 & 15.376 \\
& $RT$             & -2.783 & -2.780 & -2.646 & -2.641 & -2.638 & -2.514 & -2.239 & -2.237 & -2.131 & -3.950 & -3.947 & -3.762 \\
& $P$              & -0.672 & -0.682 & -0.630 & -0.635 & -0.643 & -0.598 & -0.153 & -0.155 & -0.149 & 0.354  & 0.360  & 0.329  \\
				
				\hline  
				\hline
			\end{tabular}   
			%	 }}
%}

\end{table*}

	\begin{table*}[htb]

	\caption{\label{table-LR2} The same as in Table \ref{table-LR1} but for $^{116}$Cd, $^{128}$Te, $^{130}$Te and $^{136}$Xe }
	\renewcommand{\arraystretch}{1.4}

			\begin{tabular}{c c c c c c c c c c c c c c}
				\hline    
				\hline  
				&&\multicolumn{3}{c}{$^{116}$Cd} &  \multicolumn{3}{c}{$^{128}$Te} &  \multicolumn{3}{c}{$^{130}$Te} &  \multicolumn{3}{c}{$^{136}$Xe} \\
				\cmidrule(lr){3-5}\cmidrule(lr){6-8}\cmidrule(lr){9-11}\cmidrule(lr){12-14}
				%	\cmidrule(lr){13-14}\cmidrule(lr){15-16}\cmidrule(lr){17-18}				
				&&AV18 & cd-Bonn & w/o &AV18 & cd-Bonn & w/o &AV18 & cd-Bonn & w/o &AV18 & cd-Bonn & w/o  \\
				
				\hline
$M_{F}$         &                  & -1.354 & -1.427 & -1.361 & -1.532 & -1.651 & -1.568 & -1.304 & -1.408 & -1.336 & -0.629 & -0.681 & -0.642 \\
\cline{3-14}
$M_{GT}$        & AA               & 3.485  & 3.773  & 3.612  & 5.169  & 5.712  & 5.464  & 4.318  & 4.790  & 4.575  & 2.450  & 2.697  & 2.571  \\
& AP               & -1.280 & -1.408 & -1.387 & -2.125 & -2.365 & -2.341 & -1.827 & -2.035 & -2.015 & -1.020 & -1.130 & -1.115 \\
& PP               & 0.425  & 0.476  & 0.477  & 0.736  & 0.832  & 0.839  & 0.639  & 0.721  & 0.729  & 0.353  & 0.397  & 0.399  \\
& MM               & 0.470  & 0.569  & 0.616  & 0.839  & 1.023  & 1.116  & 0.729  & 0.889  & 0.970  & 0.396  & 0.480  & 0.522  \\
& total            & 2.921  & 3.193  & 3.085  & 4.300  & 4.813  & 4.654  & 3.581  & 4.028  & 3.890  & 2.028  & 2.261  & 2.179  \\
\cline{3-14}
$M_{T}$         & AP               & -0.528 & -0.527 & -0.514 & -1.309 & -1.306 & -1.277 & -1.180 & -1.177 & -1.151 & -0.586 & -0.584 & -0.572 \\
& PP               & 0.195  & 0.194  & 0.189  & 0.464  & 0.463  & 0.450  & 0.418  & 0.417  & 0.406  & 0.206  & 0.205  & 0.200  \\
& MM               & -0.125 & -0.125 & -0.120 & -0.293 & -0.293 & -0.281 & -0.264 & -0.264 & -0.253 & -0.129 & -0.129 & -0.124 \\
& total            & -0.410 & -0.410 & -0.400 & -1.026 & -1.025 & -1.001 & -0.925 & -0.924 & -0.902 & -0.460 & -0.459 & -0.449 \\
\hline
$M_{\omega F}$  &                  & -1.261 & -1.329 & -1.271 & -1.521 & -1.635 & -1.559 & -1.305 & -1.404 & -1.338 & -0.625 & -0.674 & -0.640 \\
\cline{3-14}
$M_{\omega GT}$ & AA               & 3.384  & 3.659  & 3.515  & 5.211  & 5.732  & 5.506  & 4.394  & 4.849  & 4.652  & 2.431  & 2.667  & 2.555  \\
& AP               & -1.246 & -1.371 & -1.352 & -2.103 & -2.338 & -2.316 & -1.813 & -2.018 & -1.999 & -0.996 & -1.104 & -1.090 \\
& PP               & 0.414  & 0.464  & 0.466  & 0.726  & 0.820  & 0.828  & 0.631  & 0.713  & 0.720  & 0.344  & 0.387  & 0.390  \\
& MM               & 0.458  & 0.555  & 0.603  & 0.827  & 1.007  & 1.100  & 0.720  & 0.877  & 0.958  & 0.386  & 0.468  & 0.510  \\
& $\omega_+$ total & 2.836  & 3.097  & 3.004  & 4.347  & 4.838  & 4.700  & 3.658  & 4.087  & 3.966  & 2.018  & 2.240  & 2.170  \\
& $\omega_-$ total & 2.268  & 2.408  & 2.256  & 3.322  & 3.589  & 3.336  & 2.766  & 3.000  & 2.779  & 1.540  & 1.660  & 1.538  \\
\cline{3-14}
$M_{\omega T}$  & AP               & -0.521 & -0.520 & -0.507 & -1.257 & -1.254 & -1.226 & -1.133 & -1.130 & -1.105 & -0.563 & -0.561 & -0.549 \\
& PP               & 0.192  & 0.191  & 0.186  & 0.447  & 0.446  & 0.434  & 0.403  & 0.402  & 0.391  & 0.199  & 0.198  & 0.193  \\
& MM               & -0.123 & -0.123 & -0.118 & -0.284 & -0.283 & -0.271 & -0.256 & -0.256 & -0.244 & -0.126 & -0.125 & -0.120 \\
& $\omega_+$ total & -0.406 & -0.405 & -0.394 & -0.985 & -0.983 & -0.960 & -0.888 & -0.887 & -0.865 & -0.442 & -0.440 & -0.431 \\
& $\omega_-$ total & -0.252 & -0.252 & -0.248 & -0.633 & -0.632 & -0.623 & -0.571 & -0.569 & -0.563 & -0.286 & -0.285 & -0.282 \\
\hline
$M_{qF}$        &                  & -1.051 & -1.068 & -0.978 & -0.968 & -0.996 & -0.878 & -0.808 & -0.832 & -0.730 & -0.373 & -0.385 & -0.330 \\
\cline{3-14}
$M_{qGT}$       & AA               & 0.934  & 0.968  & 0.870  & 1.237  & 1.302  & 1.138  & 1.003  & 1.060  & 0.917  & 0.602  & 0.631  & 0.551  \\
& AP               & 1.804  & 1.870  & 1.714  & 2.761  & 2.885  & 2.624  & 2.334  & 2.442  & 2.216  & 1.352  & 1.409  & 1.281  \\
& PP               & -0.786 & -0.817 & -0.754 & -1.274 & -1.333 & -1.226 & -1.093 & -1.144 & -1.052 & -0.624 & -0.651 & -0.599 \\
& MM               & -1.285 & -1.389 & -1.320 & -2.229 & -2.422 & -2.311 & -1.932 & -2.100 & -2.003 & -1.069 & -1.158 & -1.103 \\
& $q_+$ total      & 2.749  & 2.881  & 2.649  & 4.106  & 4.357  & 3.968  & 3.441  & 3.659  & 3.323  & 1.992  & 2.107  & 1.916  \\
& $q_-$ total      & 1.155  & 1.159  & 1.011  & 1.342  & 1.353  & 1.103  & 1.046  & 1.055  & 0.839  & 0.667  & 0.671  & 0.549  \\
\cline{3-14}
$M_{qT}$        & AA               & 1.780  & 1.777  & 1.732  & 4.696  & 4.687  & 4.589  & 4.223  & 4.217  & 4.128  & 2.129  & 2.126  & 2.083  \\
& AP               & -1.010 & -1.008 & -0.973 & -2.346 & -2.342 & -2.265 & -2.114 & -2.110 & -2.041 & -1.038 & -1.035 & -1.003 \\
& PP               & 0.296  & 0.296  & 0.282  & 0.658  & 0.657  & 0.626  & 0.592  & 0.590  & 0.563  & 0.287  & 0.288  & 0.274  \\
& MM               & 0.111  & 0.110  & 0.100  & 0.236  & 0.239  & 0.218  & 0.213  & 0.213  & 0.193  & 0.103  & 0.104  & 0.096  \\
& $q_+$ total      & 0.998  & 0.997  & 0.978  & 2.862  & 2.855  & 2.815  & 2.569  & 2.564  & 2.530  & 1.315  & 1.314  & 1.294  \\
& $q_-$ total      & 1.136  & 1.133  & 1.103  & 3.154  & 3.151  & 3.085  & 2.834  & 2.829  & 2.770  & 1.443  & 1.442  & 1.413  \\
\hline
& $RGT$             & 5.816  & 7.303  & 8.218  & 10.822 & 13.675 & 15.488 & 9.459  & 11.953 & 13.537 & 5.183  & 6.526  & 7.364  \\
& $RT$             & -1.671 & -1.669 & -1.593 & -4.028 & -4.024 & -3.848 & -3.643 & -3.640 & -3.481 & -1.809 & -1.807 & -1.730 \\
& $P$              & 0.307  & 0.313  & 0.287  & -0.420 & -0.427 & -0.391 & -0.335 & -0.341 & -0.311 & -0.385 & -0.390 & -0.363 \\
				
				\hline  
				\hline
			\end{tabular}  
			%	 }}
	%}

\end{table*}

\begin{table}[htb]

	\caption{\label{table-ratios} The ratios between matrix elements with and without SRC.}
	\renewcommand{\arraystretch}{1.4}

			\resizebox{1.0\linewidth}{!}{ 	
			
			\begin{tabular}{c c c c c c c c c c}
				\hline    
				\hline  
           &        & $^{76}$Ge & $^{82}$Se & $^{96}$Zr & $^{100}$Mo & $^{116}$Cd & $^{128}$Te & $^{130}$Te & $^{136}$Xe \\
           \hline
				\hline
$F$           & AV18 & 0.97      & 0.98      & 0.98      & 0.98       & 0.99       & 0.98       & 0.98       & 0.98       \\
& CD-Bonn & 1.05      & 1.05      & 1.05      & 1.05       & 1.05       & 1.05       & 1.05       & 1.06       \\
$GT$          & AV18 & 0.93      & 0.93      & 0.90      & 0.91       & 0.95       & 0.92       & 0.92       & 0.93       \\
& CD-Bonn & 1.03      & 1.03      & 1.03      & 1.03       & 1.04       & 1.03       & 1.04       & 1.04       \\
$T$           & AV18 & 1.03      & 1.03      & 1.03      & 1.03       & 1.03       & 1.03       & 1.03       & 1.03       \\
& CD-Bonn & 1.03      & 1.03      & 1.03      & 1.03       & 1.02       & 1.02       & 1.02       & 1.02       \\
\hline
$\omega F$    & AV18 & 0.97      & 0.98      & 0.98      & 0.98       & 0.99       & 0.98       & 0.98       & 0.98       \\
& CD-Bonn & 1.05      & 1.05      & 1.05      & 1.04       & 1.05       & 1.05       & 1.05       & 1.05       \\
$\omega GT_+$ & AV18 & 0.93      & 0.93      & 0.91      & 0.92       & 0.94       & 0.92       & 0.92       & 0.93       \\
& CD-Bonn & 1.03      & 1.03      & 1.03      & 1.03       & 1.03       & 1.03       & 1.03       & 1.03       \\
$\omega T_+$  & AV18 & 1.03      & 1.03      & 1.03      & 1.03       & 1.03       & 1.03       & 1.03       & 1.03       \\
& CD-Bonn & 1.03      & 1.03      & 1.03      & 1.03       & 1.03       & 1.02       & 1.02       & 1.02       \\
$\omega GT_-$ & AV18 & 0.99      & 1.00      & 0.99      & 0.99       & 1.01       & 1.00       & 1.00       & 1.00       \\
& CD-Bonn & 1.07      & 1.07      & 1.08      & 1.08       & 1.07       & 1.08       & 1.08       & 1.08       \\
$\omega T_-$  & AV18 & 1.02      & 1.02      & 1.02      & 1.02       & 1.02       & 1.02       & 1.02       & 1.02       \\
& CD-Bonn & 1.02      & 1.02      & 1.02      & 1.02       & 1.02       & 1.01       & 1.01       & 1.01       \\
\hline
$qF$          & AV18 & 1.10      & 1.10      & 1.09      & 1.09       & 1.07       & 1.10       & 1.11       & 1.13       \\
& CD-Bonn & 1.13      & 1.13      & 1.12      & 1.11       & 1.09       & 1.13       & 1.14       & 1.17       \\
$qGT_+$       & AV18 & 1.03      & 1.03      & 1.03      & 1.03       & 1.04       & 1.03       & 1.04       & 1.04       \\
& CD-Bonn & 1.09      & 1.09      & 1.11      & 1.10       & 1.09       & 1.10       & 1.10       & 1.10       \\
$qT_+$        & AV18 & 1.02      & 1.02      & 1.02      & 1.02       & 1.02       & 1.02       & 1.02       & 1.02       \\
& CD-Bonn & 1.02      & 1.02      & 1.02      & 1.02       & 1.02       & 1.01       & 1.01       & 1.02       \\
$qGT_-$       & AV18 & 1.16      & 1.17      & 1.42      & 1.27       & 1.14       & 1.22       & 1.25       & 1.21       \\
& CD-Bonn & 1.17      & 1.18      & 1.45      & 1.29       & 1.15       & 1.23       & 1.26       & 1.22       \\
$qT_-$        & AV18 & 1.03      & 1.03      & 1.03      & 1.03       & 1.03       & 1.02       & 1.02       & 1.02       \\
& CD-Bonn & 1.03      & 1.03      & 1.03      & 1.03       & 1.03       & 1.02       & 1.02       & 1.02       \\
\hline
$RGT$          & AV18 & 0.70      & 0.70      & 0.69      & 0.69       & 0.71       & 0.70       & 0.70       & 0.70       \\
& CD-Bonn & 0.88      & 0.88      & 0.88      & 0.88       & 0.89       & 0.88       & 0.88       & 0.89       \\
$RT$          & AV18 & 1.05      & 1.05      & 1.05      & 1.05       & 1.05       & 1.05       & 1.05       & 1.05       \\
& CD-Bonn & 1.05      & 1.05      & 1.05      & 1.05       & 1.05       & 1.05       & 1.05       & 1.04       \\
\hline
$P$           & AV18 & 1.07      & 1.06      & 1.03      & 1.08       & 1.07       & 1.07       & 1.08       & 1.06       \\
& CD-Bonn & 1.08      & 1.07      & 1.03      & 1.09       & 1.09       & 1.09       & 1.10       & 1.07    		\\	
				\hline  
				\hline
			\end{tabular}
		}   
		\end{table}

%The nuclear matrix elements for $0\nu\beta\beta$ candidate nuclei A=76, 82, 96, 100, 116, 128, 130 and 136 are evaluated by the pn-QRPA approach systematically. 
In this work, we calculate $0\nu\beta\beta$ NMEs for eight nuclei $^{76}$Ge, $^{82}$Se, $^{96}$Zr, $^{100}$Mo, $^{116}$Cd, $^{128}$Te, $^{130}$Te and $^{136}$Xe.
The single particle levels are obtained from a Coulomb-corrected Woods-Saxon potential. In this work, we adopt a single-particle model space comprising 8 major oscillator shells (N=0$-$7) . This model space is sufficiently large to incorporate all relevant valence orbitals for the nuclei of interest and the contribution from higher oscillator shells ( N$>$7 ) is basically negligible. 

The G-matrix elements obtained from realistic NN potential (CD-Bonn) are employed to pairing interaction in BCS method for ground states and residual interaction in QRPA equation for  intermediate states. 
The BCS model is adopted to account for the residual isovector pairing interactions among nucleons (neutron-neutron and proton-proton). 
Typically, the strength of the pairing interactions $g_{pair}$ as an overall factor multiplied to the G-matrix elements is adjusted to reproduce the phenomenological pairing gaps from the five-point formula \cite{five-point} with the experimental masses taken from Ref. \cite{Wang_2021}. 
The residual interaction of the particle-particle channel of the QRPA phonon is divided into two parts as mentioned above, isoscalar and isovector parts, with corresponding renormalization factors $g_{pp}$'s. These renormalized parameters are presented in the Table \ref{gpair}, where the average value of pairing strength $\left< g_{pair}\right>$ are close to the $g_{pp}^{T=1}$ as predicted in \cite{Simkovic2013}.

In the Table \ref{table-LR1} and  \ref{table-LR2}, the $0\nu\beta\beta$-decay NMEs for eight nuclei
obtained from our pn-QRPA approach are presented. 
For each nuclei, NMEs without SRC are tabulated, as well as results with the Argonne-V18 (AV18) and CD-Bonn SRC parametrizations consistently obtained from corresponding nuclear force \cite{simkovic-SRC} are presented.
In this work, for the sake of comparison with various calculations, the factor $g_A (0)$ and $g_V(0)$ are both set to be one, of each component such as AA, AP, PP and MM, while the MM components are divided by $g_A^2$ ($g_A=1.27$) when adding to the total GT and Tensor NMEs.
%In all three cases, the same residual NN potential (CD-Bonn) is used in the calculations, the only difference of them lies in the SRC. this allows us to discuss the effect of SRC on the NMEs.

As most calculations, in the light neutrino mass mechanism, the GT parts dominate the NMEs, with the largest contribution coming from the AA components. $M_{F}$ contributes about $1/3$ from the naive analysis with Fierz rearrangement and the reduction from $M_{T}$ is about 10-20\% as in most QRPA calculations. Different behaviors for these two parts are observed in most Shell Model calculations \cite{NSM-PRL}, where smaller $M_F$ and negligible $M_T$ is observed.

%comparison with other QRPA calculations
Actually, a lot of QRPA calculations have been done for the light neutrino mass mechanism, from either the QRPA with realistic forces \cite{Simkovic2013,Suhonen-2015} or with the so-called self-consistent calculations based on the Skyrme density functional \cite{Lv2023}. Our results are close to those from the realistic force since we are using the same residual interactions, while these results are generally smaller than that of Skyme DFT, the reason still needs to be investigated though comparative studies.

While the NME differs for different many-body approaches, for different nuclei, the NME also varies. Their difference could be as large as a factor of two. Our calculations shows that $^{76}$Ge has the largest NME, almost twice larger than the smallest one from $^{136}$Xe.

Concerning the $\omega$ terms, their differences to the mass terms arise only from the energy denominator (eq.\eqref{omega}), and since $M^{0\nu}$ is less sensitive to the intermediate states excitation energies than the NME for $2\nu\beta\beta$-decay.
As a result, $M_{\omega F}$, $M_{\omega GT}$ and $M_{\omega T}$ are quite close to the corresponding terms of the mass terms. The difference for the $GT$ part is less than 10\% and for the Fermi part is less than 7\%, {\it etc}.
%Subsequently, the effect of SRC on $\omega$ terms is very similar to that of the mass terms. 
  
As observed in LSSM calculations \cite{fang2024}, the $q$ term behaves quite differently as the mass term, as different components contribute differently. In fact, as stated above, that we absorb the factor $1/3$ to the definition of $M_{qGT}$, in this sense that the q term can be written in an uniform form as sums of the Fermi, GT and Tensor parts like that for the mass and $\omega$ terms. The Fermi NME is reduced by 20-50\% compared to counterpart in the mass term. %there is about 20\%-50\% reduction in the magnitude of Fermi parts compared to the counterparts of mass terms. Combining with the enhancement of $GT$ parts, it leads to the suppression of contributions for Fermi parts to decay rates.
Meanwhile, for the GT part, the AA components in $q$ terms are largely hindered by a factor $1/3$ as explained above. Unlike the mass term, now AP components are the dominant ingredients for GT part, about twice larger in magnitude than other components. In this sense, earlier calculations \cite{Muto1989} with contributions only from AA components may give incomplete estimation over the NMEs for $qGT$ part.
The MM component from induced weak magnetic current which emerges as a sub leading order contribution according to Naive Dimension Analysis (NDA) from Chiral effective field theory, is expected to be suppressed. However, as first discovered in LSSM calculations, it appears as leading order contribution with a similar magnitude as AA or AP component, the reason of such enhancement is discussed in \cite{fang2024}. Its contribution to the total NME depends on the type of nuclear current, since there is a sign difference of weak-magnetism current for V-A and V+A currents. So the MM component comes out as an enhancement to cancel the contribution from PP component or as an cancellation to cancel the contribution from AA component. In general, compared to counterparts in mass term, all components besides AA component is getting enhanced in magnitude due to contributions from higher exchange momenta.

However, the most significant difference between the $q$ term and mass term is from the tensor part. For mass term, the tensor part comes out as a sub-leading order contribution with roughly $10\%$ corrections to the GT part. But for $q$ term, $qT$ part could be as important as $qGT$ in QRPA calculations, this is already observed in \cite{Muto1989}, and now our calculation confirms this with more components included. For $qT$ part, AA component dominates the NME, while AP component gives a reduction of about 50\% to that of AA. The PP and MM components are smaller, they give a correction about 10\%. The overall contributions of $qT$ could be as important as $qGT$.

We also find that mass term and q term are weakly correlated, given that their ratios are nearly nucleus independent, although for different nuclei their individual values are quite different.

%In addition, the different effects of SRC on $M_{qGT}$ are observed from the Table \ref{table-ratios}, in comparison to the GT parts of the mass terms. It is found that the cdb which is larger than dr by about 12\%-25\% continues to be the biggest results. The dr, rather than arg, produces the smallest NMEs, and arg is larger than dr up to 14\%-25\%. These uncertainties arised form SRC are larger than those from the mass terms. 
%This is mainly due to dominant AP terms where  
%On the other hand, the total values of the Tensor parts of $q$ terms are almost equivalent to those of the mass and $\omega$ terms, despite the presence of additional AA components that are dominant in the Tensor parts.
%While the PP and MM components are suppressed heavily compared to the corresponding one in the Tensor parts of mass terms. 
%Their sizes are small that their contributions to $M_{qT}$ are negligible. 

\begin{figure}[htbp]
	\centering
	\includegraphics[width=0.45\textwidth]{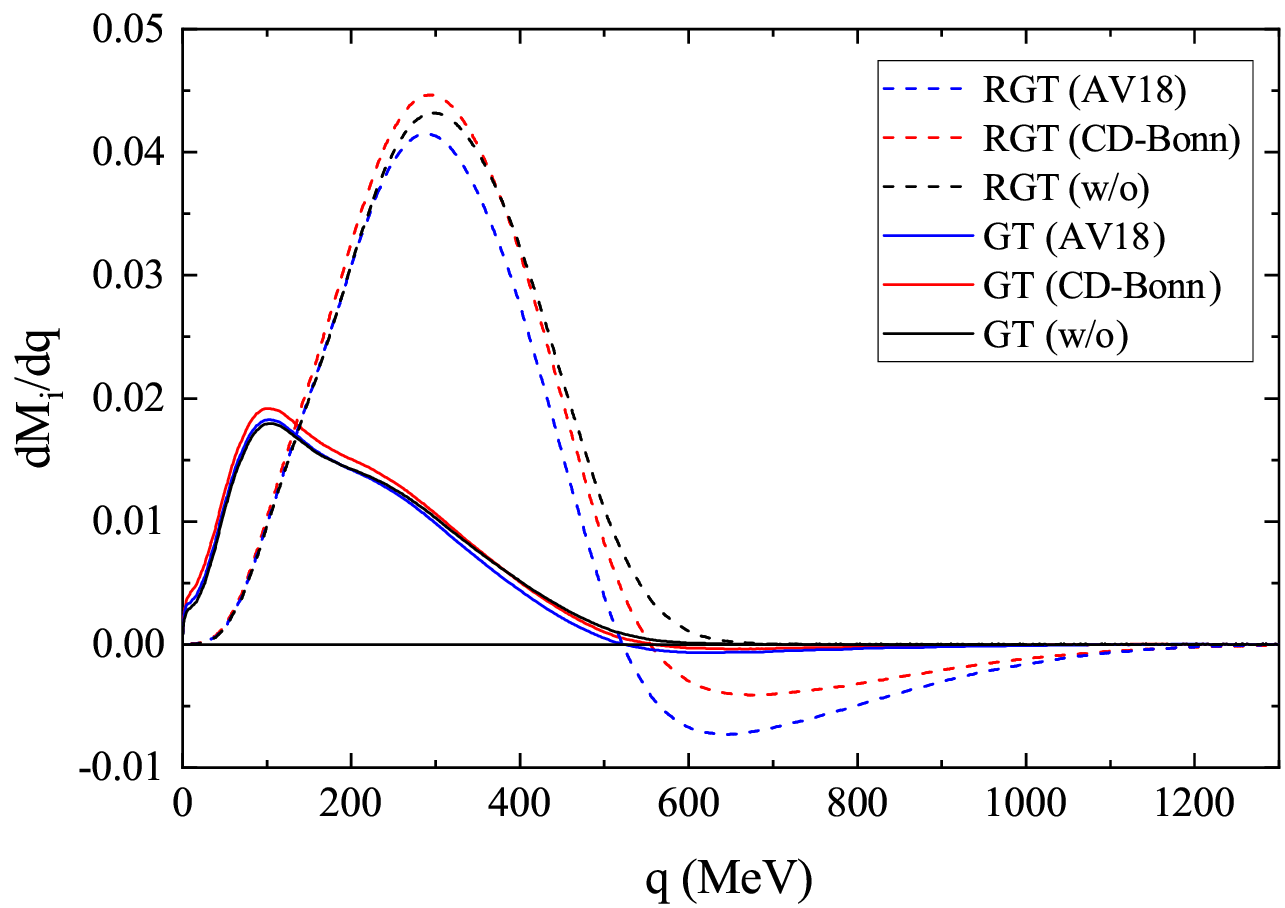}
	\caption{\label{figure:RG}(Color online) The $GT$ and $RGT$ part of NMEs for $^{76}$Ge as a function of momentum transfer $q$. `AV18' (`CD-Bonn') represents the results obtained with Argonne (CD-Bonn) SRC parametrization and `w/o' represents the results without src included.}
\end{figure}

For the R term, $M_{RGT}$ is larger than other parts such as $M_{GT}$ and $M_{qGT}$ for each nuclei and has the largest value in all LRSM NMEs.
The large values of the $M_{RGT}$ were also reported in the previous QRPA calculations  \cite{Muto1989} and LSSM calculations \cite{fang2024}. 
The same situation happens to the $RT$ parts, which also have magnitudes that are significantly larger than the corresponding parts of the mass and $\omega$ terms.

Finally for the $M_{P}$ term, it can be found that they are the smallest terms in magnitude for all LRSM NMEs and their contributions to the decay rates should be negligible. This suppresses the p-wave effect \cite{Doi:1985dx} and makes the R term dominate the $\eta$ mechanism\cite{fang2024}. The results from current calculation and previous LSSM calculation suggests that we can safely neglect this term and simplify the expression in eq.\eqref{equation:coefficients} if the limited accuracy is required.

In the last part, we focus our discussion on the effects of the src on the NMEs. As stated above, the src originates from the strong repulsive core of the nuclear force which alternates the short-range behaviors of various operators such as the two-body operator for $0\nu\beta\beta$-decay.
To understand its effect on NMEs besides the absolute values of NMEs with or without srcs presented in Table.\ref{table-LR1},\ref{table-LR2}, we give explicitly the ratios of NME with different SRC parametrizations to that without any src  
in Table.\ref{table-ratios} . %It can be found that the results for different SRC parametrizations show apparent discrepancy.   

Let's fist focus on the mass term. For $M_{GT}$, the AV18 SRC reduces the NME by about 4\%-9\% in magnitude, while the cd-Bonn SRC enhances the NME by about 4\%. The deviation between these two src is about 10\%, this agrees with other calculations\cite{simkovic-SRC}.  
For $M_{F}$, the deviation due to the different SRC is around 5\%-8\%, which is little less than the variations observed in $M_{GT}$.
For $M_{T}$, the uncertainties arising from various SRC parametrizations range from 1\% to 3\%, indicating that the effect of SRC on $M_T$ is negligible. Similar behaviors are observed for the $\omega$-term, since it is basically the same as mass term. An enhancement of the $q$-term are observed for both src parametrizations, the magnitudes are generally around 10\% for the Fermi and GT parts, but there are exceptions for the $qGT-$ part for certain nuclei, this enhancement could go as high as 40\%, this is mostly caused by the cancellations among different components. On the other hand, the important tensor part from the q term seems not affected by the src and the inclusion of src changes the NME by only a few percent. So is the P term, the general effects of src are below 10\%.

Finally, we consider the effect of SRC on the R term. While the effect of src on the tensor part is smaller as that on the tensor parts of other terms, drastic reduction is observed for the RGT part, especially for the Argonne src. The effects of src on $RGT$ have the similar but milder effects as the case of heavy neutrino mass mechanism in various literature \cite{Simkovic2013,fang2018}. While the CD-Bonn src leads to a reduction about 10\%, the Argonne src has much wilder effects, a reduction about 30\% is observed in our calculations, and also it seems that such reduction is nucleus independent. 

In order to understand the origin of this large reduction of src on the $RGT$ part, we present the transfer momentum distributions of the matrix elements $M_{RG}$ and $M_{GT}$ in Fig.\ref{figure:RG}, by comparing the results with or without src. 
One can clearly see that for $M_{GT}$, the largest contribution comes from the momentum around 100 MeV ($m_\pi$), whereas for $M_{RGT}$, major contribution is dominated by the region where momentum is around 300 MeV.  
It is mainly due to an additional $q^2$ multiplied to the spherical Bessel function $j_0 (qr)$ in the integrand (\ref{RG_equation}), which enhances the contribution from higher transfer momentum. This implies the importance of short-range contributions for this part of NME, that's why the src which alters the short-range behavior of various operator plays an important role here.
%Although the weak-magnetism current is identified as the NLO component, the high typical momentum combined with the large factor $\mu_p -\mu_n$ lead to the large magnitude of $RGT$ nuclear matrix elements.
As shown in the Fig. \ref{figure:RG}, there is a notable difference in the results of those with or without src in the region where q $>$ 600 MeV for $M_{RGT}$.
While the src is not included, we see no contribution to $M_{RGT}$ from these high momenta, similar is seen for that of $GT$ part in mass mechanism. But if the src is included, the behavior for $M_{GT}$ and $M_{RGT}$ differs greatly, a strong cancellation appears at this high momentum region for the latter. Meanwhile, a cancellation is barely observed for $M_{GT}$.
% This indicates that the large reduction (30\%) of arg comparing to the dr is mostly due to the different contributions of arg and dr in this high transfer momentum region.
% %In fact, the big difference among these three results discussed above is reasonable since that the effect of SRC mainly exist in high momentum area.  
% In fact, the large deviation associated with SRC in this region is understandable, as the influence of SRC is primarily concentrated in the high momentum area.
% The contributions from this high momentum region for the $M_{GT}$ is suppressed, hence the influences of SRC on the $M_{GT}$ are not as large as the $M_{RG}$.
Therefore, from fig.\ref{figure:RG}, we find that the reduction from src for $M_{RGT}$ mainly comes from a drastic distortion of the transition strength distributions as a function of exchange momentum. The magnitude of the changes from the two src's differs by about 20 percent. Compared to the CD-Bonn src, the AV18 src reduce the strength at low $q$ peak and enhance the high $q$ reduction. While for CD-Bonn src, the reduction at high $q$ overwhelm the enhancement at low $q$, this is in contrary to the case of $M_{GT}$. This emphasizes the importance of correctly describing the short range behaviors of the nuclear force. 

\section{CONCLUSION}
In this work, the nuclear matrix elements of $0\nu\beta\beta$ under LR symmetric model are calculated by the QRPA approach for eight nuclei: $^{76}$Ge, $^{82}$Se, $^{96}$Zr, $^{100}$Mo, $^{116}$Cd, $^{128}$Te, $^{130}$Te and $^{136}$Xe. 
The weak-magnetism components of the nucleon current are incorporated for NME calculations.
We then find that these components in the $q$ term play an important role, although they are usually considered to be suppressed. This conclusion is consistent with the previous LSSM calculation \cite{fang2024}.  
The $R$ term becomes the largest term for all LR symmetric model NMEs, hence their contributions to decay rates is also important. 
Furthermore, we discuss the effect of different SRC parametrizations on the NME. 
We find that certain parts of the NME ($e.g.$ $M_{RGT}$) are more sensitive to the SRC than other due to a large reduction from high exchange momentum region.

\section{ACKNOWLEDGEMENT}
This work is supported by National Key Research and Development Program of China (2021YFA1601300). This work is also supported by Chinese Academy of Sciences Project for Young Scientists in Basic Research (YSBR-099). The numerical calculations in this paper have been carried out on the supercomputing system in the Southern Nuclear Science Computing Center.

\nocite{*}

\bibliography{QRPA_LR}% Produces the bibliography via BibTeX.

\providecommand{\noopsort}[1]{}\providecommand{\singleletter}[1]{#1}%
\begin{thebibliography}{38}%
\makeatletter
\providecommand \@ifxundefined [1]{%
 \@ifx{#1\undefined}
}%
\providecommand \@ifnum [1]{%
 \ifnum #1\expandafter \@firstoftwo
 \else \expandafter \@secondoftwo
 \fi
}%
\providecommand \@ifx [1]{%
 \ifx #1\expandafter \@firstoftwo
 \else \expandafter \@secondoftwo
 \fi
}%
\providecommand \natexlab [1]{#1}%
\providecommand \enquote  [1]{``#1''}%
\providecommand \bibnamefont  [1]{#1}%
\providecommand \bibfnamefont [1]{#1}%
\providecommand \citenamefont [1]{#1}%
\providecommand \href@noop [0]{\@secondoftwo}%
\providecommand \href [0]{\begingroup \@sanitize@url \@href}%
\providecommand \@href[1]{\@@startlink{#1}\@@href}%
\providecommand \@@href[1]{\endgroup#1\@@endlink}%
\providecommand \@sanitize@url [0]{\catcode `\\12\catcode `\$12\catcode `\&12\catcode `\#12\catcode `\^12\catcode `\_12\catcode `\%12\relax}%
\providecommand \@@startlink[1]{}%
\providecommand \@@endlink[0]{}%
\providecommand \url  [0]{\begingroup\@sanitize@url \@url }%
\providecommand \@url [1]{\endgroup\@href {#1}{\urlprefix }}%
\providecommand \urlprefix  [0]{URL }%
\providecommand \Eprint [0]{\href }%
\providecommand \doibase [0]{https://doi.org/}%
\providecommand \selectlanguage [0]{\@gobble}%
\providecommand \bibinfo  [0]{\@secondoftwo}%
\providecommand \bibfield  [0]{\@secondoftwo}%
\providecommand \translation [1]{[#1]}%
\providecommand \BibitemOpen [0]{}%
\providecommand \bibitemStop [0]{}%
\providecommand \bibitemNoStop [0]{.\EOS\space}%
\providecommand \EOS [0]{\spacefactor3000\relax}%
\providecommand \BibitemShut  [1]{\csname bibitem#1\endcsname}%
\let\auto@bib@innerbib\@empty
%</preamble>
\bibitem [{\citenamefont {Schechter}\ and\ \citenamefont {Valle}(1982)}]{Schechter1982}%
  \BibitemOpen
  \bibfield  {author} {\bibinfo {author} {\bibfnamefont {J.}~\bibnamefont {Schechter}}\ and\ \bibinfo {author} {\bibfnamefont {J.~W.~F.}\ \bibnamefont {Valle}},\ }\href {https://doi.org/10.1103/PhysRevD.25.2951} {\bibfield  {journal} {\bibinfo  {journal} {Phys. Rev. D}\ }\textbf {\bibinfo {volume} {25}},\ \bibinfo {pages} {2951} (\bibinfo {year} {1982})}\BibitemShut {NoStop}%
\bibitem [{\citenamefont {Menéndez}(2017)}]{Menendez2018}%
  \BibitemOpen
  \bibfield  {author} {\bibinfo {author} {\bibfnamefont {J.}~\bibnamefont {Menéndez}},\ }\href {https://doi.org/10.1088/1361-6471/aa9bd4} {\bibfield  {journal} {\bibinfo  {journal} {J. Phys. G}\ }\textbf {\bibinfo {volume} {45}},\ \bibinfo {pages} {014003} (\bibinfo {year} {2017})}\BibitemShut {NoStop}%
\bibitem [{\citenamefont {Horoi}\ and\ \citenamefont {Neacsu}(2016)}]{Horoi2016}%
  \BibitemOpen
  \bibfield  {author} {\bibinfo {author} {\bibfnamefont {M.}~\bibnamefont {Horoi}}\ and\ \bibinfo {author} {\bibfnamefont {A.}~\bibnamefont {Neacsu}},\ }\href {https://doi.org/10.1103/PhysRevC.93.024308} {\bibfield  {journal} {\bibinfo  {journal} {Phys. Rev. C}\ }\textbf {\bibinfo {volume} {93}},\ \bibinfo {pages} {024308} (\bibinfo {year} {2016})}\BibitemShut {NoStop}%
\bibitem [{\citenamefont {Sen'kov}\ and\ \citenamefont {Horoi}(2013)}]{Senkov2013}%
  \BibitemOpen
  \bibfield  {author} {\bibinfo {author} {\bibfnamefont {R.~A.}\ \bibnamefont {Sen'kov}}\ and\ \bibinfo {author} {\bibfnamefont {M.}~\bibnamefont {Horoi}},\ }\href {https://doi.org/10.1103/PhysRevC.88.064312} {\bibfield  {journal} {\bibinfo  {journal} {Phys. Rev. C}\ }\textbf {\bibinfo {volume} {88}},\ \bibinfo {pages} {064312} (\bibinfo {year} {2013})}\BibitemShut {NoStop}%
\bibitem [{\citenamefont {\ifmmode~\check{S}\else \v{S}\fi{}imkovic}\ \emph {et~al.}(1999)\citenamefont {\ifmmode~\check{S}\else \v{S}\fi{}imkovic}, \citenamefont {Pantis}, \citenamefont {Vergados},\ and\ \citenamefont {Faessler}}]{Simkovic1999}%
  \BibitemOpen
  \bibfield  {author} {\bibinfo {author} {\bibfnamefont {F.}~\bibnamefont {\ifmmode~\check{S}\else \v{S}\fi{}imkovic}}, \bibinfo {author} {\bibfnamefont {G.}~\bibnamefont {Pantis}}, \bibinfo {author} {\bibfnamefont {J.~D.}\ \bibnamefont {Vergados}},\ and\ \bibinfo {author} {\bibfnamefont {A.}~\bibnamefont {Faessler}},\ }\href {https://doi.org/10.1103/PhysRevC.60.055502} {\bibfield  {journal} {\bibinfo  {journal} {Phys. Rev. C}\ }\textbf {\bibinfo {volume} {60}},\ \bibinfo {pages} {055502} (\bibinfo {year} {1999})}\BibitemShut {NoStop}%
\bibitem [{\citenamefont {Rodin}\ \emph {et~al.}(2006)\citenamefont {Rodin}, \citenamefont {Faessler}, \citenamefont {Šimkovic},\ and\ \citenamefont {Vogel}}]{Rodin2006}%
  \BibitemOpen
  \bibfield  {author} {\bibinfo {author} {\bibfnamefont {V.}~\bibnamefont {Rodin}}, \bibinfo {author} {\bibfnamefont {A.}~\bibnamefont {Faessler}}, \bibinfo {author} {\bibfnamefont {F.}~\bibnamefont {Šimkovic}},\ and\ \bibinfo {author} {\bibfnamefont {P.}~\bibnamefont {Vogel}},\ }\href {https://doi.org/https://doi.org/10.1016/j.nuclphysa.2005.12.004} {\bibfield  {journal} {\bibinfo  {journal} {Nucl. Phys. A}\ }\textbf {\bibinfo {volume} {766}},\ \bibinfo {pages} {107} (\bibinfo {year} {2006})}\BibitemShut {NoStop}%
\bibitem [{\citenamefont {Kortelainen}\ and\ \citenamefont {Suhonen}(2007)}]{Kortelainen2007}%
  \BibitemOpen
  \bibfield  {author} {\bibinfo {author} {\bibfnamefont {M.}~\bibnamefont {Kortelainen}}\ and\ \bibinfo {author} {\bibfnamefont {J.}~\bibnamefont {Suhonen}},\ }\href {https://doi.org/10.1103/PhysRevC.76.024315} {\bibfield  {journal} {\bibinfo  {journal} {Phys. Rev. C}\ }\textbf {\bibinfo {volume} {76}},\ \bibinfo {pages} {024315} (\bibinfo {year} {2007})}\BibitemShut {NoStop}%
\bibitem [{\citenamefont {\ifmmode~\check{S}\else \v{S}\fi{}imkovic}\ \emph {et~al.}(2013)\citenamefont {\ifmmode~\check{S}\else \v{S}\fi{}imkovic}, \citenamefont {Rodin}, \citenamefont {Faessler},\ and\ \citenamefont {Vogel}}]{Simkovic2013}%
  \BibitemOpen
  \bibfield  {author} {\bibinfo {author} {\bibfnamefont {F.}~\bibnamefont {\ifmmode~\check{S}\else \v{S}\fi{}imkovic}}, \bibinfo {author} {\bibfnamefont {V.}~\bibnamefont {Rodin}}, \bibinfo {author} {\bibfnamefont {A.}~\bibnamefont {Faessler}},\ and\ \bibinfo {author} {\bibfnamefont {P.}~\bibnamefont {Vogel}},\ }\href {https://doi.org/10.1103/PhysRevC.87.045501} {\bibfield  {journal} {\bibinfo  {journal} {Phys. Rev. C}\ }\textbf {\bibinfo {volume} {87}},\ \bibinfo {pages} {045501} (\bibinfo {year} {2013})}\BibitemShut {NoStop}%
\bibitem [{\citenamefont {Hyv\"arinen}\ and\ \citenamefont {Suhonen}(2015)}]{Suhonen-2015}%
  \BibitemOpen
  \bibfield  {author} {\bibinfo {author} {\bibfnamefont {J.}~\bibnamefont {Hyv\"arinen}}\ and\ \bibinfo {author} {\bibfnamefont {J.}~\bibnamefont {Suhonen}},\ }\href {https://doi.org/10.1103/PhysRevC.91.024613} {\bibfield  {journal} {\bibinfo  {journal} {Phys. Rev. C}\ }\textbf {\bibinfo {volume} {91}},\ \bibinfo {pages} {024613} (\bibinfo {year} {2015})}\BibitemShut {NoStop}%
\bibitem [{\citenamefont {Fang}\ \emph {et~al.}(2018)\citenamefont {Fang}, \citenamefont {Faessler},\ and\ \citenamefont {\ifmmode~\check{S}\else \v{S}\fi{}imkovic}}]{fang2018}%
  \BibitemOpen
  \bibfield  {author} {\bibinfo {author} {\bibfnamefont {D.-L.}\ \bibnamefont {Fang}}, \bibinfo {author} {\bibfnamefont {A.}~\bibnamefont {Faessler}},\ and\ \bibinfo {author} {\bibfnamefont {F.}~\bibnamefont {\ifmmode~\check{S}\else \v{S}\fi{}imkovic}},\ }\href {https://doi.org/10.1103/PhysRevC.97.045503} {\bibfield  {journal} {\bibinfo  {journal} {Phys. Rev. C}\ }\textbf {\bibinfo {volume} {97}},\ \bibinfo {pages} {045503} (\bibinfo {year} {2018})}\BibitemShut {NoStop}%
\bibitem [{\citenamefont {Lv}\ \emph {et~al.}(2023)\citenamefont {Lv}, \citenamefont {Niu}, \citenamefont {Fang}, \citenamefont {Yao}, \citenamefont {Bai},\ and\ \citenamefont {Meng}}]{Lv2023}%
  \BibitemOpen
  \bibfield  {author} {\bibinfo {author} {\bibfnamefont {W.-L.}\ \bibnamefont {Lv}}, \bibinfo {author} {\bibfnamefont {Y.-F.}\ \bibnamefont {Niu}}, \bibinfo {author} {\bibfnamefont {D.-L.}\ \bibnamefont {Fang}}, \bibinfo {author} {\bibfnamefont {J.-M.}\ \bibnamefont {Yao}}, \bibinfo {author} {\bibfnamefont {C.-L.}\ \bibnamefont {Bai}},\ and\ \bibinfo {author} {\bibfnamefont {J.}~\bibnamefont {Meng}},\ }\href {https://doi.org/10.1103/PhysRevC.108.L051304} {\bibfield  {journal} {\bibinfo  {journal} {Phys. Rev. C}\ }\textbf {\bibinfo {volume} {108}},\ \bibinfo {pages} {L051304} (\bibinfo {year} {2023})}\BibitemShut {NoStop}%
\bibitem [{\citenamefont {Barea}\ \emph {et~al.}(2015)\citenamefont {Barea}, \citenamefont {Kotila},\ and\ \citenamefont {Iachello}}]{Barea2015}%
  \BibitemOpen
  \bibfield  {author} {\bibinfo {author} {\bibfnamefont {J.}~\bibnamefont {Barea}}, \bibinfo {author} {\bibfnamefont {J.}~\bibnamefont {Kotila}},\ and\ \bibinfo {author} {\bibfnamefont {F.}~\bibnamefont {Iachello}},\ }\href {https://doi.org/10.1103/PhysRevC.91.034304} {\bibfield  {journal} {\bibinfo  {journal} {Phys. Rev. C}\ }\textbf {\bibinfo {volume} {91}},\ \bibinfo {pages} {034304} (\bibinfo {year} {2015})}\BibitemShut {NoStop}%
\bibitem [{\citenamefont {Deppisch}\ \emph {et~al.}(2020)\citenamefont {Deppisch}, \citenamefont {Graf}, \citenamefont {Iachello},\ and\ \citenamefont {Kotila}}]{Deppisch2020}%
  \BibitemOpen
  \bibfield  {author} {\bibinfo {author} {\bibfnamefont {F.~F.}\ \bibnamefont {Deppisch}}, \bibinfo {author} {\bibfnamefont {L.}~\bibnamefont {Graf}}, \bibinfo {author} {\bibfnamefont {F.}~\bibnamefont {Iachello}},\ and\ \bibinfo {author} {\bibfnamefont {J.}~\bibnamefont {Kotila}},\ }\href {https://doi.org/10.1103/PhysRevD.102.095016} {\bibfield  {journal} {\bibinfo  {journal} {Phys. Rev. D}\ }\textbf {\bibinfo {volume} {102}},\ \bibinfo {pages} {095016} (\bibinfo {year} {2020})}\BibitemShut {NoStop}%
\bibitem [{\citenamefont {Rodr\'{\i}guez}\ and\ \citenamefont {Mart\'{\i}nez-Pinedo}(2010)}]{Rodriguez2010}%
  \BibitemOpen
  \bibfield  {author} {\bibinfo {author} {\bibfnamefont {T.~R.}\ \bibnamefont {Rodr\'{\i}guez}}\ and\ \bibinfo {author} {\bibfnamefont {G.}~\bibnamefont {Mart\'{\i}nez-Pinedo}},\ }\href {https://doi.org/10.1103/PhysRevLett.105.252503} {\bibfield  {journal} {\bibinfo  {journal} {Phys. Rev. Lett.}\ }\textbf {\bibinfo {volume} {105}},\ \bibinfo {pages} {252503} (\bibinfo {year} {2010})}\BibitemShut {NoStop}%
\bibitem [{\citenamefont {Song}\ \emph {et~al.}(2014)\citenamefont {Song}, \citenamefont {Yao}, \citenamefont {Ring},\ and\ \citenamefont {Meng}}]{Song2014}%
  \BibitemOpen
  \bibfield  {author} {\bibinfo {author} {\bibfnamefont {L.~S.}\ \bibnamefont {Song}}, \bibinfo {author} {\bibfnamefont {J.~M.}\ \bibnamefont {Yao}}, \bibinfo {author} {\bibfnamefont {P.}~\bibnamefont {Ring}},\ and\ \bibinfo {author} {\bibfnamefont {J.}~\bibnamefont {Meng}},\ }\href {https://doi.org/10.1103/PhysRevC.90.054309} {\bibfield  {journal} {\bibinfo  {journal} {Phys. Rev. C}\ }\textbf {\bibinfo {volume} {90}},\ \bibinfo {pages} {054309} (\bibinfo {year} {2014})}\BibitemShut {NoStop}%
\bibitem [{\citenamefont {Song}\ \emph {et~al.}(2017)\citenamefont {Song}, \citenamefont {Yao}, \citenamefont {Ring},\ and\ \citenamefont {Meng}}]{Song2017}%
  \BibitemOpen
  \bibfield  {author} {\bibinfo {author} {\bibfnamefont {L.~S.}\ \bibnamefont {Song}}, \bibinfo {author} {\bibfnamefont {J.~M.}\ \bibnamefont {Yao}}, \bibinfo {author} {\bibfnamefont {P.}~\bibnamefont {Ring}},\ and\ \bibinfo {author} {\bibfnamefont {J.}~\bibnamefont {Meng}},\ }\href {https://doi.org/10.1103/PhysRevC.95.024305} {\bibfield  {journal} {\bibinfo  {journal} {Phys. Rev. C}\ }\textbf {\bibinfo {volume} {95}},\ \bibinfo {pages} {024305} (\bibinfo {year} {2017})}\BibitemShut {NoStop}%
\bibitem [{\citenamefont {Yao}\ \emph {et~al.}(2015)\citenamefont {Yao}, \citenamefont {Song}, \citenamefont {Hagino}, \citenamefont {Ring},\ and\ \citenamefont {Meng}}]{Yao2015}%
  \BibitemOpen
  \bibfield  {author} {\bibinfo {author} {\bibfnamefont {J.~M.}\ \bibnamefont {Yao}}, \bibinfo {author} {\bibfnamefont {L.~S.}\ \bibnamefont {Song}}, \bibinfo {author} {\bibfnamefont {K.}~\bibnamefont {Hagino}}, \bibinfo {author} {\bibfnamefont {P.}~\bibnamefont {Ring}},\ and\ \bibinfo {author} {\bibfnamefont {J.}~\bibnamefont {Meng}},\ }\href {https://doi.org/10.1103/PhysRevC.91.024316} {\bibfield  {journal} {\bibinfo  {journal} {Phys. Rev. C}\ }\textbf {\bibinfo {volume} {91}},\ \bibinfo {pages} {024316} (\bibinfo {year} {2015})}\BibitemShut {NoStop}%
\bibitem [{\citenamefont {Yao}\ \emph {et~al.}(2022)\citenamefont {Yao}, \citenamefont {Meng}, \citenamefont {Niu},\ and\ \citenamefont {Ring}}]{Yao2022}%
  \BibitemOpen
  \bibfield  {author} {\bibinfo {author} {\bibfnamefont {J.~M.}\ \bibnamefont {Yao}}, \bibinfo {author} {\bibfnamefont {J.}~\bibnamefont {Meng}}, \bibinfo {author} {\bibfnamefont {Y.~F.}\ \bibnamefont {Niu}},\ and\ \bibinfo {author} {\bibfnamefont {P.}~\bibnamefont {Ring}},\ }\href {https://doi.org/https://doi.org/10.1016/j.ppnp.2022.103965} {\bibfield  {journal} {\bibinfo  {journal} {Prog. Part. Nucl. Phys.}\ }\textbf {\bibinfo {volume} {126}},\ \bibinfo {pages} {103965} (\bibinfo {year} {2022})}\BibitemShut {NoStop}%
\bibitem [{\citenamefont {Agostini}\ \emph {et~al.}(2023)\citenamefont {Agostini}, \citenamefont {Benato}, \citenamefont {Detwiler}, \citenamefont {Men\'endez},\ and\ \citenamefont {Vissani}}]{Matteo2023}%
  \BibitemOpen
  \bibfield  {author} {\bibinfo {author} {\bibfnamefont {M.}~\bibnamefont {Agostini}}, \bibinfo {author} {\bibfnamefont {G.}~\bibnamefont {Benato}}, \bibinfo {author} {\bibfnamefont {J.~A.}\ \bibnamefont {Detwiler}}, \bibinfo {author} {\bibfnamefont {J.}~\bibnamefont {Men\'endez}},\ and\ \bibinfo {author} {\bibfnamefont {F.}~\bibnamefont {Vissani}},\ }\href {https://doi.org/10.1103/RevModPhys.95.025002} {\bibfield  {journal} {\bibinfo  {journal} {Rev. Mod. Phys.}\ }\textbf {\bibinfo {volume} {95}},\ \bibinfo {pages} {025002} (\bibinfo {year} {2023})}\BibitemShut {NoStop}%
\bibitem [{\citenamefont {Yao}\ \emph {et~al.}(2020)\citenamefont {Yao}, \citenamefont {Bally}, \citenamefont {Engel}, \citenamefont {Wirth}, \citenamefont {Rodr\'{\i}guez},\ and\ \citenamefont {Hergert}}]{Yao2020}%
  \BibitemOpen
  \bibfield  {author} {\bibinfo {author} {\bibfnamefont {J.~M.}\ \bibnamefont {Yao}}, \bibinfo {author} {\bibfnamefont {B.}~\bibnamefont {Bally}}, \bibinfo {author} {\bibfnamefont {J.}~\bibnamefont {Engel}}, \bibinfo {author} {\bibfnamefont {R.}~\bibnamefont {Wirth}}, \bibinfo {author} {\bibfnamefont {T.~R.}\ \bibnamefont {Rodr\'{\i}guez}},\ and\ \bibinfo {author} {\bibfnamefont {H.}~\bibnamefont {Hergert}},\ }\href {https://doi.org/10.1103/PhysRevLett.124.232501} {\bibfield  {journal} {\bibinfo  {journal} {Phys. Rev. Lett.}\ }\textbf {\bibinfo {volume} {124}},\ \bibinfo {pages} {232501} (\bibinfo {year} {2020})}\BibitemShut {NoStop}%
\bibitem [{\citenamefont {Caurier}\ \emph {et~al.}(1996)\citenamefont {Caurier}, \citenamefont {Nowacki}, \citenamefont {Poves},\ and\ \citenamefont {Retamosa}}]{NSM-PRL}%
  \BibitemOpen
  \bibfield  {author} {\bibinfo {author} {\bibfnamefont {E.}~\bibnamefont {Caurier}}, \bibinfo {author} {\bibfnamefont {F.}~\bibnamefont {Nowacki}}, \bibinfo {author} {\bibfnamefont {A.}~\bibnamefont {Poves}},\ and\ \bibinfo {author} {\bibfnamefont {J.}~\bibnamefont {Retamosa}},\ }\href {https://doi.org/10.1103/PhysRevLett.77.1954} {\bibfield  {journal} {\bibinfo  {journal} {Phys. Rev. Lett.}\ }\textbf {\bibinfo {volume} {77}},\ \bibinfo {pages} {1954} (\bibinfo {year} {1996})}\BibitemShut {NoStop}%
\bibitem [{\citenamefont {Muto}\ \emph {et~al.}(1989)\citenamefont {Muto}, \citenamefont {Bender},\ and\ \citenamefont {Klapdor}}]{Muto1989}%
  \BibitemOpen
  \bibfield  {author} {\bibinfo {author} {\bibfnamefont {K.}~\bibnamefont {Muto}}, \bibinfo {author} {\bibfnamefont {E.}~\bibnamefont {Bender}},\ and\ \bibinfo {author} {\bibfnamefont {H.~V.}\ \bibnamefont {Klapdor}},\ }\href {https://doi.org/10.1007/BF01294219} {\bibfield  {journal} {\bibinfo  {journal} {Z. Phys. A}\ }\textbf {\bibinfo {volume} {334}},\ \bibinfo {pages} {187} (\bibinfo {year} {1989})}\BibitemShut {NoStop}%
\bibitem [{\citenamefont {Pantis}\ \emph {et~al.}(1996)\citenamefont {Pantis}, \citenamefont {\ifmmode~\check{S}\else \v{S}\fi{}imkovic}, \citenamefont {Vergados},\ and\ \citenamefont {Faessler}}]{Pantis1996}%
  \BibitemOpen
  \bibfield  {author} {\bibinfo {author} {\bibfnamefont {G.}~\bibnamefont {Pantis}}, \bibinfo {author} {\bibfnamefont {F.}~\bibnamefont {\ifmmode~\check{S}\else \v{S}\fi{}imkovic}}, \bibinfo {author} {\bibfnamefont {J.~D.}\ \bibnamefont {Vergados}},\ and\ \bibinfo {author} {\bibfnamefont {A.}~\bibnamefont {Faessler}},\ }\href {https://doi.org/10.1103/PhysRevC.53.695} {\bibfield  {journal} {\bibinfo  {journal} {Phys. Rev. C}\ }\textbf {\bibinfo {volume} {53}},\ \bibinfo {pages} {695} (\bibinfo {year} {1996})}\BibitemShut {NoStop}%
\bibitem [{\citenamefont {Suhonen}\ and\ \citenamefont {Civitarese}(1998)}]{SUHONEN1998}%
  \BibitemOpen
  \bibfield  {author} {\bibinfo {author} {\bibfnamefont {J.}~\bibnamefont {Suhonen}}\ and\ \bibinfo {author} {\bibfnamefont {O.}~\bibnamefont {Civitarese}},\ }\href {https://doi.org/https://doi.org/10.1016/S0370-1573(97)00087-2} {\bibfield  {journal} {\bibinfo  {journal} {Phys. Rep.}\ }\textbf {\bibinfo {volume} {300}},\ \bibinfo {pages} {123} (\bibinfo {year} {1998})}\BibitemShut {NoStop}%
\bibitem [{\citenamefont {Tomoda}(1991)}]{Tomoda1991}%
  \BibitemOpen
  \bibfield  {author} {\bibinfo {author} {\bibfnamefont {T.}~\bibnamefont {Tomoda}},\ }\href {https://doi.org/10.1088/0034-4885/54/1/002} {\bibfield  {journal} {\bibinfo  {journal} {Rep. Prog. Phys.}\ }\textbf {\bibinfo {volume} {54}},\ \bibinfo {pages} {53} (\bibinfo {year} {1991})}\BibitemShut {NoStop}%
\bibitem [{\citenamefont {Sarkar}\ \emph {et~al.}(2020)\citenamefont {Sarkar}, \citenamefont {Iwata},\ and\ \citenamefont {Raina}}]{Sarkar2020}%
  \BibitemOpen
  \bibfield  {author} {\bibinfo {author} {\bibfnamefont {S.}~\bibnamefont {Sarkar}}, \bibinfo {author} {\bibfnamefont {Y.}~\bibnamefont {Iwata}},\ and\ \bibinfo {author} {\bibfnamefont {P.~K.}\ \bibnamefont {Raina}},\ }\href {https://doi.org/10.1103/PhysRevC.102.034317} {\bibfield  {journal} {\bibinfo  {journal} {Phys. Rev. C}\ }\textbf {\bibinfo {volume} {102}},\ \bibinfo {pages} {034317} (\bibinfo {year} {2020})}\BibitemShut {NoStop}%
\bibitem [{\citenamefont {Iwata}\ and\ \citenamefont {Sarkar}(2021)}]{Iwata2021}%
  \BibitemOpen
  \bibfield  {author} {\bibinfo {author} {\bibfnamefont {Y.}~\bibnamefont {Iwata}}\ and\ \bibinfo {author} {\bibfnamefont {S.}~\bibnamefont {Sarkar}},\ }\href {https://www.frontiersin.org/articles/10.3389/fspas.2021.727880} {\bibfield  {journal} {\bibinfo  {journal} {Front. Astron. Space Sci.}\ }\textbf {\bibinfo {volume} {8}},\ \bibinfo {pages} {727880} (\bibinfo {year} {2021})}\BibitemShut {NoStop}%
\bibitem [{\citenamefont {Šimkovic}\ \emph {et~al.}(2017)\citenamefont {Šimkovic}, \citenamefont {Štefánik},\ and\ \citenamefont {Dvornický}}]{Simkovic2017}%
  \BibitemOpen
  \bibfield  {author} {\bibinfo {author} {\bibfnamefont {F.}~\bibnamefont {Šimkovic}}, \bibinfo {author} {\bibfnamefont {D.}~\bibnamefont {Štefánik}},\ and\ \bibinfo {author} {\bibfnamefont {R.}~\bibnamefont {Dvornický}},\ }\href {https://www.frontiersin.org/articles/10.3389/fphy.2017.00057} {\bibfield  {journal} {\bibinfo  {journal} {Front. Phys.}\ }\textbf {\bibinfo {volume} {5}},\ \bibinfo {pages} {57} (\bibinfo {year} {2017})}\BibitemShut {NoStop}%
\bibitem [{\citenamefont {Štefánik}\ \emph {et~al.}(2015)\citenamefont {Štefánik}, \citenamefont {Dvornický}, \citenamefont {Šimkovic},\ and\ \citenamefont {Vogel}}]{Reexam-LR}%
  \BibitemOpen
  \bibfield  {author} {\bibinfo {author} {\bibfnamefont {D.}~\bibnamefont {Štefánik}}, \bibinfo {author} {\bibfnamefont {R.}~\bibnamefont {Dvornický}}, \bibinfo {author} {\bibfnamefont {F.}~\bibnamefont {Šimkovic}},\ and\ \bibinfo {author} {\bibfnamefont {P.}~\bibnamefont {Vogel}},\ }\href {https://doi.org/10.1103/PhysRevC.92.055502} {\bibfield  {journal} {\bibinfo  {journal} {Phys. Rev. C}\ }\textbf {\bibinfo {volume} {92}},\ \bibinfo {pages} {055502} (\bibinfo {year} {2015})}\BibitemShut {NoStop}%
\bibitem [{\citenamefont {Fang}\ \emph {et~al.}(2024)\citenamefont {Fang}, \citenamefont {Brown},\ and\ \citenamefont {\ifmmode~\check{S}\else \v{S}\fi{}imkovic}}]{fang2024}%
  \BibitemOpen
  \bibfield  {author} {\bibinfo {author} {\bibfnamefont {D.-L.}\ \bibnamefont {Fang}}, \bibinfo {author} {\bibfnamefont {B.~A.}\ \bibnamefont {Brown}},\ and\ \bibinfo {author} {\bibfnamefont {F.}~\bibnamefont {\ifmmode~\check{S}\else \v{S}\fi{}imkovic}},\ }\href {https://doi.org/10.1103/PhysRevC.110.045502} {\bibfield  {journal} {\bibinfo  {journal} {Phys. Rev. C}\ }\textbf {\bibinfo {volume} {110}},\ \bibinfo {pages} {045502} (\bibinfo {year} {2024})}\BibitemShut {NoStop}%
\bibitem [{\citenamefont {Doi}\ \emph {et~al.}(1985)\citenamefont {Doi}, \citenamefont {Kotani},\ and\ \citenamefont {Takasugi}}]{Doi:1985dx}%
  \BibitemOpen
  \bibfield  {author} {\bibinfo {author} {\bibfnamefont {M.}~\bibnamefont {Doi}}, \bibinfo {author} {\bibfnamefont {T.}~\bibnamefont {Kotani}},\ and\ \bibinfo {author} {\bibfnamefont {E.}~\bibnamefont {Takasugi}},\ }\href {https://doi.org/10.1143/PTPS.83.1} {\bibfield  {journal} {\bibinfo  {journal} {Prog. Theor. Phys. Suppl.}\ }\textbf {\bibinfo {volume} {83}},\ \bibinfo {pages} {1} (\bibinfo {year} {1985})}\BibitemShut {NoStop}%
\bibitem [{\citenamefont {Xing}(2012)}]{Xing:2011ur}%
  \BibitemOpen
  \bibfield  {author} {\bibinfo {author} {\bibfnamefont {Z.-z.}\ \bibnamefont {Xing}},\ }\href {https://doi.org/10.1103/PhysRevD.85.013008} {\bibfield  {journal} {\bibinfo  {journal} {Phys. Rev. D}\ }\textbf {\bibinfo {volume} {85}},\ \bibinfo {pages} {013008} (\bibinfo {year} {2012})}\BibitemShut {NoStop}%
\bibitem [{\citenamefont {Kortelainen}\ \emph {et~al.}(2007)\citenamefont {Kortelainen}, \citenamefont {Civitarese}, \citenamefont {Suhonen},\ and\ \citenamefont {Toivanen}}]{Kortelainen2007-2}%
  \BibitemOpen
  \bibfield  {author} {\bibinfo {author} {\bibfnamefont {M.}~\bibnamefont {Kortelainen}}, \bibinfo {author} {\bibfnamefont {O.}~\bibnamefont {Civitarese}}, \bibinfo {author} {\bibfnamefont {J.}~\bibnamefont {Suhonen}},\ and\ \bibinfo {author} {\bibfnamefont {J.}~\bibnamefont {Toivanen}},\ }\href {https://doi.org/https://doi.org/10.1016/j.physletb.2007.01.054} {\bibfield  {journal} {\bibinfo  {journal} {Phys. Lett. B}\ }\textbf {\bibinfo {volume} {647}},\ \bibinfo {pages} {128} (\bibinfo {year} {2007})}\BibitemShut {NoStop}%
\bibitem [{\citenamefont {Miller}\ and\ \citenamefont {Spencer}(1976)}]{Gerald1976}%
  \BibitemOpen
  \bibfield  {author} {\bibinfo {author} {\bibfnamefont {G.~A.}\ \bibnamefont {Miller}}\ and\ \bibinfo {author} {\bibfnamefont {J.~E.}\ \bibnamefont {Spencer}},\ }\href {https://doi.org/https://doi.org/10.1016/0003-4916(76)90073-7} {\bibfield  {journal} {\bibinfo  {journal} {Ann. Phys.}\ }\textbf {\bibinfo {volume} {100}},\ \bibinfo {pages} {562} (\bibinfo {year} {1976})}\BibitemShut {NoStop}%
\bibitem [{\citenamefont {\ifmmode~\check{S}\else \v{S}\fi{}imkovic}\ \emph {et~al.}(2009)\citenamefont {\ifmmode~\check{S}\else \v{S}\fi{}imkovic}, \citenamefont {Faessler}, \citenamefont {M\"uther}, \citenamefont {Rodin},\ and\ \citenamefont {Stauf}}]{simkovic-SRC}%
  \BibitemOpen
  \bibfield  {author} {\bibinfo {author} {\bibfnamefont {F.}~\bibnamefont {\ifmmode~\check{S}\else \v{S}\fi{}imkovic}}, \bibinfo {author} {\bibfnamefont {A.}~\bibnamefont {Faessler}}, \bibinfo {author} {\bibfnamefont {H.}~\bibnamefont {M\"uther}}, \bibinfo {author} {\bibfnamefont {V.}~\bibnamefont {Rodin}},\ and\ \bibinfo {author} {\bibfnamefont {M.}~\bibnamefont {Stauf}},\ }\href {https://doi.org/10.1103/PhysRevC.79.055501} {\bibfield  {journal} {\bibinfo  {journal} {Phys. Rev. C}\ }\textbf {\bibinfo {volume} {79}},\ \bibinfo {pages} {055501} (\bibinfo {year} {2009})}\BibitemShut {NoStop}%
\bibitem [{\citenamefont {Rodin}\ and\ \citenamefont {Faessler}(2011)}]{Rodin2011}%
  \BibitemOpen
  \bibfield  {author} {\bibinfo {author} {\bibfnamefont {V.}~\bibnamefont {Rodin}}\ and\ \bibinfo {author} {\bibfnamefont {A.}~\bibnamefont {Faessler}},\ }\href {https://doi.org/10.1103/PhysRevC.84.014322} {\bibfield  {journal} {\bibinfo  {journal} {Phys. Rev. C}\ }\textbf {\bibinfo {volume} {84}},\ \bibinfo {pages} {014322} (\bibinfo {year} {2011})}\BibitemShut {NoStop}%
\bibitem [{\citenamefont {Bender}\ \emph {et~al.}(2000)\citenamefont {Bender}, \citenamefont {Rutz}, \citenamefont {Reinhard},\ and\ \citenamefont {Maruhn}}]{five-point}%
  \BibitemOpen
  \bibfield  {author} {\bibinfo {author} {\bibfnamefont {M.}~\bibnamefont {Bender}}, \bibinfo {author} {\bibfnamefont {K.}~\bibnamefont {Rutz}}, \bibinfo {author} {\bibfnamefont {P.~G.}\ \bibnamefont {Reinhard}},\ and\ \bibinfo {author} {\bibfnamefont {J.~A.}\ \bibnamefont {Maruhn}},\ }\href {https://doi.org/10.1007/s10050-000-4504-z} {\bibfield  {journal} {\bibinfo  {journal} {Eur. Phys. J. A}\ }\textbf {\bibinfo {volume} {8}},\ \bibinfo {pages} {59} (\bibinfo {year} {2000})}\BibitemShut {NoStop}%
\bibitem [{\citenamefont {Wang}\ \emph {et~al.}(2021)\citenamefont {Wang}, \citenamefont {Huang}, \citenamefont {Kondev}, \citenamefont {Audi},\ and\ \citenamefont {Naimi}}]{Wang_2021}%
  \BibitemOpen
  \bibfield  {author} {\bibinfo {author} {\bibfnamefont {M.}~\bibnamefont {Wang}}, \bibinfo {author} {\bibfnamefont {W.}~\bibnamefont {Huang}}, \bibinfo {author} {\bibfnamefont {F.}~\bibnamefont {Kondev}}, \bibinfo {author} {\bibfnamefont {G.}~\bibnamefont {Audi}},\ and\ \bibinfo {author} {\bibfnamefont {S.}~\bibnamefont {Naimi}},\ }\href {https://doi.org/10.1088/1674-1137/abddaf} {\bibfield  {journal} {\bibinfo  {journal} {Chin. Phys. C}\ }\textbf {\bibinfo {volume} {45}},\ \bibinfo {pages} {030003} (\bibinfo {year} {2021})}\BibitemShut {NoStop}%
\end{thebibliography}%

\end{document}